 \documentclass[12pt, draftclsnofoot, onecolumn]{IEEEtran}

\renewcommand\IEEEkeywordsname{Index Terms}

\newcommand\tab[1][1cm]{\hspace*{#1}}
\usepackage{xcolor}
\usepackage{adjustbox}
\usepackage{graphicx}
\usepackage[cmex10]{amsmath}
\usepackage{dsfont}
\usepackage{cite}
\usepackage{lipsum}
\usepackage{tabu}
\usepackage{bm}
\newcommand{\myMatrix}[1]{\bm{\mathit{#1}}}
\usepackage{xparse}
\usepackage{amsfonts}
\usepackage{amssymb}
\usepackage{textcomp}
\usepackage{resizegather}
\usepackage{multirow}
\usepackage{epsfig}
\usepackage{floatrow}
\usepackage{caption}
\usepackage{subcaption}
\usepackage{latexsym}

\usepackage{algorithm,algpseudocode}

\makeatletter
\renewcommand{\fnum@algorithm}{\fname@algorithm}
\makeatother

\usepackage{epstopdf}
\usepackage{amsmath, amsthm, amssymb,enumitem}
\usepackage{verbatim}
\graphicspath{{./figures/}}
\usepackage{graphicx}
\usepackage{booktabs}
\renewcommand\thesubsection{\Alph{subsection}}
\usepackage{color}
\usepackage{adjustbox}
\newtheorem{theorem}{Theorem}

\let\svthefootnote\thefootnote

\author{Ahmed Roushdy,
        Abolfazl Seyed Motahari, Mohammed Nafie and \\ Deniz G\"{u}nd\"{u}z
        \thanks{A. Roushdy is with the Department of Engineering Mathematics and Physics, Faculty of Engineering and the Intelligent Networks Center (WINC), Nile University (email: ahmed.elkordy17@imperial.ac.uk). A. S. Motahari is with the Department of Computer Engineering, Sharif University of Technology (email: motahari@sharif.edu). M. Nafie is with the Electronics and Communications Department, Cairo University (email: mnafie@ieee.org). D. G\"{u}nd\"{u}z is with the Information Processing and Communication Lab, Imperial College London (email: d.gunduz@imperial.ac.uk).}
        \thanks{The material in this paper was presented in part at the IEEE Wireless Communications and Networking Conference (WCNC) in Barcelona, Spain, in April 2018.}
        }
        
\begin{document}
\title{Cache-Aided    Combination Networks \\ with Interference}

 \maketitle
\begin{abstract}
Centralized coded caching  and delivery is studied 
for a radio access combination network (RACN), whereby a set of $H$  edge nodes (ENs), connected to a cloud server via orthogonal fronthaul links with limited capacity,  serve a total  of $K$  user equipments (UEs) over wireless  links. The cloud   server is assumed to hold a library of $N$ files, each of size $F$ bits; and each user,  equipped with a cache
of size $\mu_R N F$ bits, is connected to a distinct set of $r$  ENs    each of which equipped with a cache  of size $\mu_T N F$ bits,   where $\mu_T$,  $\mu_R \in [0,1]$ are the fractional cache capacities of the UEs and the ENs, respectively. The objective is  to minimize the normalized delivery time (NDT), which refers to the worst case delivery latency when each user requests a single    distinct  file from the library.      Three   coded caching and transmission  schemes are  considered, namely  the\textit{ MDS-IA}, \textit{soft-transfer} and \textit{ zero-forcing (ZF)} schemes.  MDS-IA    utilizes   maximum distance separable (MDS) codes in the  placement phase and real interference alignment (IA) in the delivery phase. The  achievable NDT for this scheme  is presented for $r=2$  and      arbitrary  fractional cache sizes $\mu_T$ and $\mu_R$, and also for arbitrary value of  $r$ and  fractional  cache size $\mu_T$ when the cache capacity of the UE is above a certain threshold. The soft-transfer scheme utilizes soft-transfer of coded symbols to ENs that implement  ZF over the edge links.  The  achievable NDT for this scheme  is presented   for arbitrary $r$ and arbitrary     fractional cache sizes  $\mu_T$ and $\mu_R$. The  last scheme utilizes   ZF  between the ENs and the UEs    without  the  participation  of  the  cloud  server  in  the delivery phase. The  achievable NDT for this scheme  is presented  for an   arbitrary value of  $r$   when the total cache size at a pair of  UE and EN is sufficient to store the whole library, i.e., $\mu_T+\mu_R \geq 1$. The results indicate that  the fronthaul   capacity  determines which scheme  achieves a  better performance in terms of the  NDT,   and the soft-transfer scheme becomes favorable as the fronthaul capacity increases.  
\end{abstract}

\section{Introduction}
Proactively caching  is  considered a promising solution for the growing  network traffic and latency   for  future communication networks \cite{GolrezaeiFemtocaching, GregoryDtoD, BAstug:ICC:14, Blasco:ISIT:14,  Somuyiwa:JSAC:18}. A centralized coded proactive caching scheme was introduced  in \cite{maddah2014},  and it is shown to provide significant coding gains with respect to  classical uncoded caching. Decentralized coded caching
is considered  in \cite{Maddah-Ali:2015:DCC:2824005.2824006, 7999228}, where each user randomly stores some bits from each file independently
of  the other users. More recently,  coded caching has been extended to  wireless radio access  networks (RANs), where transmitters and/or receivers are equipped with cache memories.  Cache-aided delivery over a noisy broadcast channel is considered in  \cite{7996417 } and \cite{8359316}. Cache-aided delivery from multiple transmitters is considered in \cite{7282567, 2016arXiv160500203X
,2017arXiv170304349P,sengupta2016cloud,koh2017cloud,girgis2017decentralized, 8422193,e19070366 }. It is shown in \cite{7282567} that caches at the transmitters can improve the sum degrees of freedom (DoF) by allowing cooperation among transmitters for interference mitigation. In \cite{ 2016arXiv160500203X } and \cite{7857805} this model is extended  to  an interference network with  $K_T$ transmitters and $K_R$   receivers, where both the transmitters  and
receivers are equipped with cache memories. An achievable scheme exploiting real interference alignment (IA) for the general $K_T\times K_R$ network is proposed in \cite{2017arXiv170304349P}, which also considers decentralized caching at the users.  An interference  network with random topology is considered in \cite{Mital:WCNC:18}.

While the above works assume that the transmitter caches are large enough to store all the database, the fog-aided RAN (F-RAN) model \cite{sengupta2016cloud} allows the delivery of contents from the cloud server to the edge-nodes (ENs) through dedicated fronthaul links. Coded caching for the F-RAN scenario with cache-enabled ENs is studied in \cite{sengupta2016cloud}. The authors propose a centralized  coded caching scheme to minimize  the normalized delivery time (NDT), which measures the worst case delivery latency with respect to an interference-free baseline system in the high
signal-to-noise ratio (SNR) regime. In \cite{koh2017cloud}, the authors consider a wireless fronthaul that enables coded multicasting. In \cite{girgis2017decentralized}, decentralized coded caching is studied for a RAN architecture with two ENs, in which both the ENs and the users have caches. In \cite{8422193}, this model is extended to an arbitrary number of  ENs and users.  We note that the models in \cite{sengupta2016cloud, koh2017cloud, girgis2017decentralized,8422193 }  assume a fully connected interference network between the ENs and users. A partially connected  RAN is studied in \cite{e19070366} from an online caching perspective.

If each EN is connected  to a subset of the users through dedicated error free orthogonal links, the corresponding architecture is known as a \textit{combination network}.  Coded caching in a combination network is studied in \cite{ji2015fundamental, tang2016coded, zewail2017coded}. In such networks, the server is connected to a set of
$H$ relay nodes, which communicate to $K={H \choose r}$ users, such that each user is connected to a distinct set of $r$ relay nodes, where $r$ is refered to  as the \textit{receiver connectivity}. The links are assumed to be error- and interference-free.  The objective is to determine the minimax link load, defined as the minimum achievable value of  the maximum load among all the links (proportional to the download time) and  over all possible 
 demand combinations. Note that, although the delivery from the ENs to the users takes place over orthogonal links,  that is, there are no multicasting  opportunities as  in\cite{maddah2014}, the fact that the messages for multiple users are delivered from the server to each relay through a single link allows coded delivery to offer   gains similarly to \cite{maddah2014}. The authors of \cite{tang2016coded}  consider a class of combination networks 
that satisfy the resolvability property, which require $H$ to be  divisible
by $r$. A combination network in which both the relays and the users are equipped with caches is  presented  in \cite{zewail2017coded}. For the case when there are no caches at the relays, the authors are able   to achieve the same performance as in \cite{tang2016coded} without requiring the  resolvability property. 
 
In this paper we study the centralized caching problem in a RACN  with cache-enabled user equipments (UEs) and ENs,  as depicted  in Fig.  \ref{Fig. 1}. Our work differs  from the aforementioned prior works \cite{sengupta2016cloud, koh2017cloud, girgis2017decentralized,8422193 } as  we consider a partially connected interference channel from the ENs to the UEs, instead of a fully connected RAN architecture. This may be due to physical constraints that block the signals or the long distance between some of the EN-UE pairs. The  network   from the server to the UEs, where ENs act as relays for the UEs they serve, is similar to the combination network architecture  \cite{ji2015fundamental, tang2016coded, zewail2017coded}; however, we consider interfering wireless links  from the ENs to the UEs  instead of dedicated links, and study the normalized delivery time in the high SNR regime. The authors in \cite{2017arXiv170809117X} study the NDT for  a partially connected $(K+L-1)\times K$ interference channel with caches at both the transmitters  and the receivers, where each  receiver is  connected   to $L$ consecutive  transmitters. Our work is different from \cite{2017arXiv170809117X}, since we  take into consideration the fronthaul links from the server to the ENs, and consider a network topology in which  the number of transmitters (ENs in our model) is less than or equal to the number of receivers, which we believe is a more realistic scenario. 

We formulate the minimum NDT problem for a given   \textit{receiver connectivity} $r$.  Then, we propose three  centralized caching and delivery schemes;  in particular, the \textit{MDS-IA} scheme that we proposed in our previous work \cite{8377403},  the  \textit{soft-transfer} scheme and the  \textit{zero-forcing (ZF)} scheme.  The MDS-IA scheme   exploits real IA to minimize the NDT
for receiver connectivity of $ r = 2$. We then extend this scheme
to an arbitrary receiver connectivity of  $r$ assuming a certain cache
capacity    at the UEs while an   arbitrary cache capacity at the ENs. For this scheme, we show that increasing the
receiver connectivity  for the same number of ENs and UEs
will decrease the  NDT for the specific cache capacity
region studied at the UEs, while the reduction in the   NDT depends on
the fronthaul capacity. On the other, in the soft-transfer scheme the server delivers quantized channel input symbols to the ENs in order to enable them to implement ZF transmission to the UEs to minimize the NDT
for an arbitrary receiver connectivity   and  cache capacity at   both the ENs and the UEs.  The ZF scheme is presented  when the total cache size at  one   UE and one  EN is sufficient
to store the  entire library,  i.e., $\mu_T+\mu_R\geq 1$, then  the  cloud server can remain silent during the delivery phase and all users’ requests can
be satisfied by  ZF  from the ENs to the  UEs.

Our results show that  the  best scheme in terms of the NDT  depends on the  fronthaul capacity   and  the cache sizes.  For the case when the total  cache size of the EN and UE is not   sufficient to store the  entire library, i.e., $\mu_T+\mu_R<1$,  the MDS-IA scheme  achieves a  smaller NDT when the fronthaul capacity is relatively  limited,  while  the soft-transfer scheme performs better as the fronthaul capacity increases.     On the other hand,  when the total  cache size of the EN and UE is  sufficient to store the entire library,   the ZF scheme achieves  a smaller NDT than the other proposed schemes   when the fronthaul capacity is relatively  limited.  

The rest of the paper is organized as follows. In Section \ref{system_model}, we introduce the system model and the  performance measure. In Section  \ref{MR}, the main results of the paper are presented. The MDS-IA scheme is  presented in Section \ref{Scheme1}, while the soft-transfer scheme is introduced in Section \ref{Scheme}. After that,  the ZF scheme is presented in Section \ref{zero-forcing}, while the numerical results are  presented in  section \ref{Numerical Results}. 
Finally, the paper is concluded in Section \ref{Conclusion}. 
 
\subsection{Notation}
We denote sets with calligraphic symbols and vectors with bold symbols. The set of integers $\{1,\ldots, N\}$ is denoted by $[N]$. The cardinality of set $\mathcal{A}$ is denoted by $|\mathcal{A}|$. We use  the function $(x)^+$ to return  $ \mathbf{max}$($x,0$).

\section{System Model and Performance Measure} \label{system_model} 
 \begin{figure}
\centering
\includegraphics[width=8cm,height=8cm,keepaspectratio]{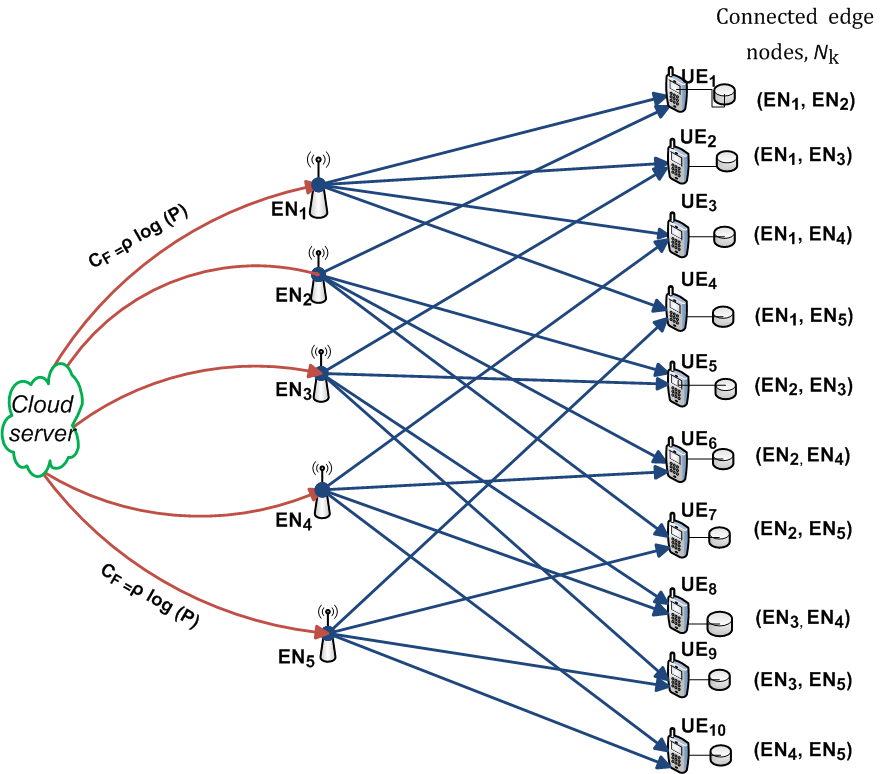}
\caption{RACN  architecture with receiver connectivity $r=2$, where $H=5$ ENs serve $K = 10$ UEs.}
\label{Fig. 1}
\end{figure}
\subsection{System Model}
We consider the  $H\times K$ RACN  architecture as illustrated in  Fig. \ref{Fig. 1}, which consists of a cloud server and   a set of $H$ ENs, $\mathcal{E}\stackrel{\Delta}{=}\{\mathrm{EN}_1,\ldots,\mathrm{EN}_H\}$, that help the cloud server  to serve the requests from a set of $K$  UEs, $\mathcal{U}\stackrel{\Delta}{=}\{\mathrm{UE}_1,\ldots,\mathrm{UE}_K\}$.   The cloud is connected to each ENs via orthogonal fronthaul links of capacity $C_F$ bits per symbol, where the symbol refers to a single use of the edge channel from the ENs to the UEs. The edge network from the ENs to the users is a partially connected interference channel, where $\mathrm{UE}_k\in\mathcal{U}$ is connected to a distinct set of $r$ ENs, where $r < H$ 
is referred to as the $\textit{receiver connectivity}$. The number of UEs is $K={H \choose r}$, which means that  $H\leq K$. In  this  architecture, $\mathrm{EN}_i$, $i\in[H]$,  is connected to $L={H-1\choose r-1}=\frac{rK}{H}$  UEs.

The cloud server  holds a library of $N$ files, $\mathcal{W}\stackrel{\Delta}{=}\{W_{1},\ldots,W_{N}$\}, each of size $F$ bits. We assume that the UEs request files from this library only. Each  UE is equipped with a cache memory  of size $\mu_R NF$ bits,    while each  EN  is equipped with a cache memory of size $\mu_T N F$,  where $\mu_T$,  $\mu_R \in [0,1]$, are the fractional cache capacities of the UEs and the ENs, respectively. We define two  parameters, $t_U=\mu_RK$ and $t_E=\mu_RL$, where the former is the normalized cache capacity (per file) available across all the UEs, while the latter is the normalized cache capacity of the UEs  connected to a particular edge node.  
We denote the set of  UEs connected to $\mathrm{EN}_i$ by $\mathcal{K}_i$, where $|\mathcal{K}_i|=L$, and the set of  ENs connected to  $\mathrm{UE}_k$ by $\mathcal{N}_k$, where  $|\mathcal{N}_k|=r$. We will use the function $\mathrm {Index}(i, k)$ : $[H]\times [K]$ $\,\to\, [L] \cup \epsilon$,  which returns $\epsilon $ if $\mathrm{UE}_k$ is not served by $\mathrm{EN}_i$, and otherwise returns   the relative order of  $\mathrm{UE}_k$  among the $L$ UEs served by $\mathrm{EN}_i$    with the assumption that the  $L$ UEs in $\mathcal{K}_i$ are sorted in ascending order. For example, in Fig.  \ref{Fig. 1}, we have $\mathcal{K}_1=\{1, 2, 3, 4\}$, $\mathcal{K}_3=\{2, 5, 8, 9\}$ and
\begin{align*}
\mathrm {Index}(1, 2)&=2, \; \; \; \; \; \mathrm {Index}(1,3)=3, \; \; \; \; \mathrm{Index}(1,5)=\epsilon,\\
\mathrm {Index}(3, 2)&=1, \; \; \; \; \;  \mathrm{Index}(3,5)=2,  \; \; \; \; \mathrm{Index}(3,1)=\epsilon.
\end{align*}

The system operates in two phases: a \textit{ placement phase} and a \textit{delivery phase}. The placement phase takes place when the
traffic load is low, and  the network nodes  are given access to the entire library $\mathcal{W}$. $\mathrm{UE}_k$, $k\in[K]$,    and    $\mathrm{EN}_i$, $i\in[H]$,  are  then able to fill
 their  caches using the library without any prior knowledge of the future 
demands or the  channel coefficients.     Let $Z_k$ and $U_i$, $k\in[K]$, $i\in[H]$, denote the cache contents of $\mathrm{UE}_k$  and $\mathrm{EN}_i$  at the end of the placement phase, respectively.  We consider centralized placement; that is, the cache contents of UEs and the ENs, donated by  $Z_1,  \dots , Z_k$, are coordinated jointly.  

In the delivery phase, $\mathrm{UE}_k$, $k\in[K]$, requests  file $W_{d_k}$ from the library, $d_k \in [N]$.  We define $\mathbf{d}=[d_1,. . . , d_K ] \in [N]^{K}$ as the demand vector. Once the demands are received, the cloud server sends message $\mathbf{G}_i=(G_i(t))_{t=1}^{T_F}$ of blocklength $T_F$ to $\mathrm{EN}_i$, $i\in[H]$, via the
fronthaul link. This message is limited to $T_FC_F$ bits to guarantee correct decoding at $\mathrm{EN}_i$ with high probability.
 In this paper, we consider half-duplex ENs; that is, ENs start transmitting only after receiving their messages from the cloud server. This is called \textit{serial transmission} in  \cite{sengupta2016cloud}, and the overall latency is the sum of the  latencies in the fronthaul and the edge connections.  $\mathrm{EN}_i$ has an encoding function  that maps     the cache contents $U_i$, fronthaul message $\mathbf{G}_i$, the demand
vector $\mathbf{d}$, and the channel coefficients $ \mathbf{H} \overset{\Delta}{=}\{h_{k,i}\}_{k\in[K], i\in [H]}$, where $h_{k,i}$ denotes the complex channel gain from $\mathrm{EN}_i$ to $\mathrm{UE}_k$,  to a channel input vector $\mathbf{V}_i=(V_i(t))_{t=1}^{T_E}$ of blocklength $T_E$, which must satisfy  an average  power constraint of $\mathbf{P}$, i.e., $E\big[\frac{1}{T_E} \mathbf{ V}_i \mathbf{V}_i^\intercal \big] \leq \mathbf{P}$. $\mathrm{UE}_k$ decodes its requested file as  $\hat{W}_{d_k}$ by using its  cache contents $Z_k$, the received signal $\mathbf{Y}_k=(Y_k(t))_{t=1}^{T_E}$, as well as its knowledge of the channel gain  matrix $\mathbf{H}$ and  the demand vector $\mathbf{d}$.  We have
\begin{equation}
Y_k(t)=\sum_{i \in \mathcal{N}_k} h_{k,i} V_i(t)+n_k(t), 
\end{equation}
where $n_k(t) \sim \mathcal{C N}(0, 1)$ denotes the  independent additive  complex Gaussian noise
at the $k$th user.  The channel gains are independent and identically distributed (i.i.d.) according to a continuous  distribution, and remain constant within each transmission interval. Similarly to \cite{7282567, 2016arXiv160500203X
,2017arXiv170304349P,sengupta2016cloud,koh2017cloud }, we assume that perfect channel state information is available at all the terminals of network. The probability of error for  a coding scheme, consisting of the  cache placement, cloud
encoding, EN encoding, and user decoding functions,  is defined as 
\begin{equation}
P_e=\max_{\mathbf{d} \in [N]^{K}} \max_{k \in [K]} P_e( \hat{W}_{d_k} \neq W_{d_k}),
\end{equation}
which is the worst-case probability of error over all possible
demand vectors  and  all the  users. We say that a coding scheme is
$\textit{feasible}$,  if we have $P_e \,\to\, $ 0 when $F\,\to\,  \infty$, for almost all  realizations of the channel matrix $\mathbf{H}$.

\setlength{\textfloatsep}{1cm}
\subsection{Performance Measure}

We will consider\textit{ the normalized delivery time (NDT)} in the high SNR regime \cite{7864374}  as the  performance measure. Note that the capacity of the edge network scales with the SNR. Hence, to make sure that the fronthaul  links do not constitute a bottleneck, we let   $C_F = \rho \log P$, where $\rho$ is called the  \textit{fronthaul multiplexing gain}.  The multiplexing gain  is the pre-log term in the capacity expression \cite{4418479, 4252335}, and   an important indicator  of  the capacity behaviour in the high SNR regime. For given  $\mu_T$, $\mu_R$ and fronthaul  multiplexing gain $\rho$, we say that  $\delta (\mu_R, \mu_T, \rho)$ is an \textit{achievable} NDT if there exists a sequence of feasible codes that satisfy 
\begin{equation}
 \delta (\mu_R, \mu_T, \rho) = \lim_{P, F\to \infty}\sup{ \frac{(T_F+T_E)\log P } {F}}.
\end{equation}
We additionally define the fronthaul NDT as 
\begin{equation}
 \delta_F (\mu_R, \mu_T, \rho)  = \lim_{P, F\to \infty}\sup{ \frac{T_F\log P } {F}}, 
\end{equation}
and the edge NDT as 
\begin{equation}
 \delta_E (\mu_R, \mu_T, \rho) = \lim_{P, F\to \infty}\sup{ \frac{T_E\log P } {F}}, 
\end{equation}
such that the end-to-end NDT is the sum of the fronthaul and edge NDTs. 
We define the minimum NDT for a
given  $(\mu_R, \mu_T, \rho)$ tuple as
\begin{equation*}
\delta^\star  (\mu_R, \mu_T, \rho)=\inf \{ \delta  (\mu_R, \mu_T, \rho)) : \delta  (\mu_R, \mu_T, \rho) \text{ is achievable}\}.
\end{equation*}

\section{Main Result} \label{MR}
 The main results of the paper are  stated in the following   theorems.
 \begin{theorem} For an  $H\times K$  RACN  architecture,   with  fractional   cache capacities  of $\mu_R$ and $\mu_T$, fronthaul multiplexing gain  $\rho \geq 0$, number of files $N\geq K$, and considering centralized cache placement, the following NDT is achievable by the MDS-IA scheme  for  integer values of $t_E$:
\begin{equation}
\label{ax}
 \delta_{\text{MDS-IA}} (\mu_R, \mu_T, \rho)= \frac{L-t_E}{r}\left[ \frac{r-1}{L} +  \frac{1}{t_E+1}\left(1+\frac{   (1-\mu_T r)^+}{\rho} \right) \right]
 \end{equation}
for a receiver connectivity of $r=2$, or for arbitrary receiver connectivity when $t_E \geq L-2$. 
\end{theorem}
\begin{theorem}
For the same  RACN  architecture, the  following NDT is achievable by  the soft-transfer scheme  for  integer values of  $t_U $
\begin{equation}
\delta_{\text{soft}} (\mu_R, \mu_T, \rho) =(K-t_U) \left[\frac{1}{\min \{H+t_U, K\}}+\frac{ (1-\mu_T)}{ H \rho} \right].
\end{equation}
\end{theorem}

 \begin{theorem}
For the same RACN  architecture with $\mu_R + \mu_T \geq 1$ the  following NDT is achievable by  the ZF  scheme  for  integer values of  $t_R $ where $t_R=\frac{(\mu_R+\mu_T-1)K}{\mu_T }$:
\begin{equation}
\delta_{\text{ZF}} (\mu_R, \mu_T, \rho) = \left(\frac{K-t_R}{\min \{H+t_R, K\}} \right) \mu_T.
\end{equation}
\end{theorem}

\textbf{Remark 1.}  \textit{The achievable NDT $\delta(\mu_T,\mu_R, \rho)$ is a convex function of $\mu_T$ and  $\mu_R$  for every
value of $\rho\geq 0$ \cite{sengupta2016cloud}. For any two $(\mu_T^1, \mu_R^1)$ and $(\mu_T^2, \mu_R^2)$ pairs, convex combination of the corresponding achievable  NDT values can be achieved through memory and time sharing. This would require dividing each of the files in the library into two parts, which have the normalized cache capacities as specified in these pairs. Then the delivery schemes specified for these two achievable points are used sequentially in a time-division manner. Hence, for a given $\mu_T$,  when $t_E$ is not an integer for the MDS-IA scheme, or $t_U$ is not integer for the soft-transfer scheme, or $t_R$ is not integer for the ZF scheme, we can write $\mu_R=\alpha \mu_R^1+(1-\alpha)\mu_R^2$ for some $\alpha \in[0, 1]$,  where $\mu_R^1$ and $\mu_R^2$ are two values that lead to integer normalized cache capacities $t_E$, $t_U$, or $t_E$, with $\mu_R^1>\mu_R^2$. By  applying  memory time-sharing  as in \cite{sengupta2016cloud}, the following  NDT is achievable
 \begin{equation}
 \label{3}
 \delta(\mu_R, \mu_T, \rho)|_{\text{M-sharing}}= \alpha \delta( \mu_R^1, \mu_T, \rho)+(1-\alpha) \delta(\mu_R^2, \mu_T, \rho).
\end{equation}}
\begin{figure}
\centering
\includegraphics[width=8cm,height=8cm,keepaspectratio]{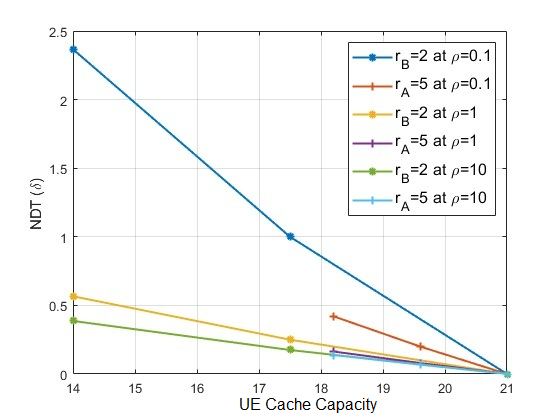}
\caption{Comparison of the  achievable NDT for a $7\times21$ RACN  architecture  with library $N=21$ files  for different receiver connectivity  and fronthaul multiplexing gains when there is no cache memory at the ENs.}
  \label{xx}
\end{figure}
\textbf{Remark 2.}   \textit{  For the same  RACN  architecture with  $\mu_R + \mu_T \geq 1$ and $\rho\leq \rho_\text{th}$, where  
\begin{equation}
\rho_\text{th}\triangleq \frac{  (1-\mu_T r)^+ \left(\delta_2 + \frac{(1-\alpha)}{\alpha} \delta_1 \right)}{ \left(\frac{K}{\min \{H, K\}} \right) \frac{\mu_T r}{\alpha } -\delta_2 \left((r-1)(\mu_R^2+\frac{1}{L})+1 \right)-\delta_1 (\frac{1}{\alpha}-1)  \left((r-1)(\mu_R^1+\frac{1}{L})+1 \right)},
\end{equation}
 with $\delta_i=\frac{1-\mu_R^i}{\mu_R^i+\frac{1}{L}}$, for $i=1,2$, while $\mu_R^1$, $\mu_R^2$ and $\alpha$ can be calculated  using memory sharing, the ZF scheme achieves a smaller NDT than the other schemes. This is due to the fact that when the fronthaul multiplexing gain is small, it is better to avoid using the fronthaul links.}
 
\textbf{Remark 3.} \textit{ From Theorem 1  Eqn. \eqref{ax},  when $r\geq  2 $, the  NDT achieved by  the MDS-IA scheme  is given by
\begin{equation*}
\label{bb}
\delta _{\text{MDS-IA}}(\mu_R, \mu_T, \rho) =\begin{cases}\frac{2}{ r} \left(\frac{r-1}{L}+ \frac{1}{L-1} \left(1+\frac{(1-\mu_T r)^+}{\rho} \right) \right), & t_E=L-2\\\ \frac{1}{L} \left( 1+ \frac{(1-\mu_T r)^+}{\rho r}\right), &  t_E=L-1\end{cases}. 
\end{equation*}
 Consider two different RACN  architectures with $H$ ENs, denoted by RACN-A and RACN-B, with receiver connectivities $r_A$ and $r_B$, respectively, where $r_A + r_B = H$ and $r_A\geq r_B$. The two networks have the same number of UEs $K= \binom {r_A+r_B} {r_A}=\binom {r_A+r_B} {r_B}$, but the number of UEs each EN connects to is different, and is given by  $L_x=\frac{K}{H}r_x$, $x\in\{A,B\}$.     We illustrate the achievable NDT performance of  the MDS-IA scheme in a $7\times 21$ RACN     in Fig. \ref{xx}   setting $r_A=5$ and $r_B=2$   with no cahce memory at the ENs  for  different fronthaul multiplexing gains. We observe from the figure that,   with the same UE  cache  capacity   the achievable NDT of network RACN-A is less than or equal to  that of network RACN-B, and the gap between the  two increases as the fronthaul multiplexing gain  decreases.   This suggests  that the increased connectivity  helps in reducing the NDT despite  potentially increasing the interference as well, and  the gap between the two achievable NDTs  for  RACN-A and RAN-B becomes negligible as the fronthaul multiplexing gain increases, i.e., $\rho\rightarrow \infty$.}

\section{ MDS-IA Scheme}\label{Scheme1}
  We first present the MDS-IA scheme without cache memories at the ENs. Afterwards, we
will extend the results to the case with cache memories at the ENs.

\let\thefootnote\svthefootnote

\subsection{MDS-IA Scheme without  Cache Memories at the ENs}\label{MDS-IA no cache}
\subsubsection{ \textbf{Cache Placement Phase}} \label{Cache Placement Phase No Cache}
We use the  placement scheme proposed in \cite{zewail2017coded}, where the cloud server divides each file into  $r$ equal-size non-overlapping subfiles.  Then, it encodes the  subfiles using an $(H, r)$ maximum distance separable (MDS) code \cite{Lin:2004:ECC:983680}. The resulting coded chunks, each of size $F/r$ bits,  are denoted by $f_n^i$, where $n$ is the file index and $i\in[H]$ is the  index of the coded chunk. 
 $\mathrm{EN}_i$  will act  as an edge  server for the  encoded chunk $f_n^i$, $ i \in [H]$. Note that, thanks to the MDS code, any $r$ encoded chunks are sufficient to reconstruct a  file.
 
Each encoded chunk $f_n^i$ is  further divided into ${L\choose t_E}$ equal-size non-overlapping pieces, each of  which is denoted by $f^i_{n, \mathcal{T}}$, where  $\mathcal{T}\subseteq[L]$, $|\mathcal{T}|=t_E$. The pieces  $f^i_{n, \mathcal{T}}$, $\forall n$, are stored in the cache memory of  $\mathrm{UE}_k$ if $k\in  \mathcal{K}_i$ and $\mathrm{Index}(i, k)\in \mathcal{T}$; that is, the pieces of chunk $i$, $i\in[H]$, are stored by the $L$ UEs connected to $\mathrm{EN}_i$. At the end of the  placement phase, each user stores
$Nr\binom {L-1} {t_E-1}$ pieces, each of size $\frac{F}{r{L\choose t_E}}$
bits, which sum up to $\mu_R NF$ bits,  satisfying the memory constraint with equality. We will next illustrate  the placement phase through an  example.

\textbf{Example 1.}
 $\;$ Consider the   RACN  depicted in Fig.  \ref{Fig. 1} , where $H=5$, $K=N=10$,  $r=2$ and $L=4$.  The cloud server   divides each file  into $r=2$ subfiles. These subfiles are then  encoded  using a $(5, 2)$ MDS code. As a result, there are $5$ coded chunks, denoted by  $f_n^i$, $n\in[10]$, $i\in[5]$, each of size $F/2$ bits. For $t_E=1$, i.e., $\mu_R=1/L$, each encoded chunk  $f_n^i$ is further divided into $\binom {L} {t_E}=4$ pieces $f^i_{n, \mathcal{T}}$, where $\mathcal{T}\subseteq[4]$ and $|\mathcal{T}|=t_E=1$. Cache contents of each user are listed in TABLE \ref{Table:a}. Observe that each user stores two 
pieces of the encoded chunks of each file for a total of 10 files, i.e., $\frac{5}{2}F$ bits, which
satisfies the memory constraint.

\begin{table*}[]
  \centering
\captionsetup{justification=raggedright,singlelinecheck=false}
\begin{adjustbox}{width=1\textwidth}
\small
    \begin{tabular}{|l|l|l|l|l|l|l|l|l|l|l|}
        \hline
        User & ~ $\mathrm{UE}_1$                                    & ~   $\mathrm{UE}_2$                                  & ~                                       $\mathrm{UE}_3$ & ~                 $\mathrm{UE}_4$                   & ~         $\mathrm{UE}_5$                             & ~     $\mathrm{UE}_6$ & ~ $\mathrm{UE}_7$ & ~    $\mathrm{UE}_8$                                  & ~  $\mathrm{UE}_9$                                   & ~                                             $\mathrm{UE}_{10}$ \\ \hline
        Cache Contents   & $f_{n,1}^1$, $f_{n,1}^2$  & $f_{n,2}^1$, $f_{n,1}^3$   & $f_{n,3}^1$, $f_{n,1}^4$     & $f_{n,4}^1$, $f_{n,1}^5$   & $f_{n,2}^2$, $f_{n,2}^3$    & $f_{n,3}^2$, $f_{n,2}^4$   & $f_{n,4}^2$, $f_{n,2}^5$  & $f_{n,3}^3$, $f_{n,3}^4$    & $f_{n,4}^3$, $f_{n,3}^5$   & $f_{n,4}^4$, $f_{n,4}^5$           \\
        \hline
    \end{tabular}
  \caption{Cache contents after the placement phase for the RACN  scenario considered in Example 1, where $K=N=10$, $r=2$, $L=4$,  $t_E=1$ and $\mu_R=\frac{1}{4}$. }
  \label{Table:a}
 \end{adjustbox}
\end{table*}

\begin{table*}[]
\centering
\captionsetup{justification=raggedright,singlelinecheck=false}
\begin{adjustbox}{width=.9\textwidth}
\small
\begin{tabular}{|l|l|l|l|l|}
\hline
\multicolumn{1}{|c|}{$\mathrm{EN}_1$}                                     & \multicolumn{1}{c|}{$\mathrm{EN}_2$}                 & \multicolumn{1}{c|}{$\mathrm{EN}_3$}                & \multicolumn{1}{c|}{$\mathrm{EN}_4$}                 & \multicolumn{1}{c|}{$\mathrm{EN}_5$}                 \\ \hline
\multicolumn{1}{|c|}{$\mathbf{X}_{1}^{1,2}=f_{1,2}^1+f_{2,1}^1$} & $\mathbf{X}_{2}^{1,2}=f_{1,2}^2+f_{5,1}^2$ & $\mathbf{X}_{3}^{1,2}=f_{2,2}^3+f_{5,1}^3$ & $\mathbf{X}_{4}^{1,2}=f_{3,2}^4+f_{6,1}^4$  & $\mathbf{X}_{5}^{1,2}=f_{4,2}^5+f_{7,1}^5$  \\
$\mathbf{X}_{1}^{1,3}=f_{1,3}^1+f_{3,1}^1$                       & $\mathbf{X}_{2}^{1,3}=f_{1,3}^2+f_{6,1}^2$  & $\mathbf{X}_{3}^{1,3}=f_{2,3}^3+f_{8,1}^3$ & $\mathbf{X}_{4}^{1,3}=f_{3,3}^4+f_{8,1}^4$  & $\mathbf{X}_{5}^{1,3}=f_{4,3}^5+f_{9,1}^5$  \\
$\mathbf{X}_{1}^{1,4}=f_{1,4}^1+f_{4,1}^1$                       & $\mathbf{X}_{2}^{1,4}=f_{1,4}^2+f_{7,1}^2$  & $\mathbf{X}_{3}^{1,4}=f_{2,4}^3+f_{9,1}^3$ & $\mathbf{X}_{4}^{1,4}=f_{3,4}^4+f_{10,1}^4$ & $\mathbf{X}_{5}^{1,4}=f_{4,4}^5+f_{10,1}^5$ \\
$\mathbf{X}_{1}^{2,3}=f_{2,3}^1+f_{3,2}^1$                       & $\mathbf{X}_{2}^{2,3}=f_{5,3}^2+f_{6,2}^2$  & $\mathbf{X}_{3}^{2,3}=f_{5,3}^3+f_{8,2}^3$ & $\mathbf{X}_{4}^{2,3}=f_{6,3}^1+f_{8,2}^4$  & $\mathbf{X}_{5}^{2,3}=f_{7,3}^5+f_{9,2}^5$  \\
$\mathbf{X}_{1}^{2,4}=f_{2,4}^1+f_{4,2}^1$                       & $\mathbf{X}_{2}^{2,4}=f_{5,4}^2+f_{7,2}^2$  & $\mathbf{X}_{3}^{2,4}=f_{5,4}^3+f_{9,2}^3$ & $\mathbf{X}_{4}^{2,4}=f_{6,4}^1+f_{10,2}^4$ & $\mathbf{X}_{5}^{2,4}=f_{7,4}^5+f_{10,2}^5$ \\
$\mathbf{X}_{1}^{3,4}=f_{3,4}^1+f_{4,3}^1$                       & $\mathbf{X}_{2}^{3,4}=f_{6,4}^2+f_{7,3}^2$  & $\mathbf{X}_{3}^{3,4}=f_{8,4}^3+f_{9,3}^3$ & $\mathbf{X}_{4}^{3,4}=f_{8,4}^1+f_{10,3}^4$ & $\mathbf{X}_{5}^{3,4}=f_{9,4}^5+f_{10,3}^5$ \\ \hline
\end{tabular}
\caption{The  data delivered  from the cloud server  to each EN for Example 1.}
\label{Table:b}
\end{adjustbox}
\end{table*}
\subsubsection{\textbf{Delivery Phase}}\label{Delivery phase no cache}
The delivery phase is carried out in two steps. The first step  is the delivery from the  cloud server to the ENs, and the second step is the delivery from the ENs to the UEs.

 $\textbf{Step 1}$:  Delivery from the  cloud server to the  ENs
For each  $(t_E+1)$-element subset $\mathcal{S}$ of $[L]$, i.e., $\mathcal{S}\subseteq[L]$ and $|\mathcal{S}|=t_E+1$, the cloud server will deliver the following message to $\mathrm{EN}_i$:
\begin{equation}\label{coded-multicasting}
\textbf{X}_i^\mathcal{S} \triangleq \bigoplus_{k: k\in \mathcal{K}_i, Index(i, k)\in \mathcal{S}} f^i_{{d_k}, \mathcal{S} \backslash \mathrm{Index}(i, k)}.
\end{equation}
Overall, for  given  $\mathbf{d}$, the following set of messages will be delivered  to $EN_i$
\begin{equation}
\{\textbf{X}_i^\mathcal{S}: \mathcal{S}\subseteq[L], |\mathcal{S}|=t_E+1\},
\end{equation}
which makes a total of  $ {L\choose t_E+1} \frac{F}{r{L\choose t_E}}$ bits.
 The fronthaul  NDT from the cloud server to the ENs   is then given by 
\begin{equation}
 \label{Fron} 
 \delta_{F}(\mu_R, \mu_T, \rho) =\frac{    \binom {L} {t_E+1} }{r{L\choose t_E} \rho} = \frac{L-t_E}{(t_E+1)r\rho}.
\end{equation}
The message to be delivered to each  EN in Example 1 is given in TABLE \ref{Table:b}, and we have $\delta_{F}(\frac{1}{4},0,\rho)=\frac{3}{4\rho}$.

\begin{algorithm}[h] \footnotesize
  Algorithm 1: Generator for $\mathds{A}$, $\mathds{B}$ and $\mathds{C}$ Matrices
  \begin{algorithmic}[1]
\State $\mathds{A}=[\;]$,   $\mathds{B}=[\;]$,  $\mathds{C}=[\;]$, $g=0$ 
 \State FOR  $k=1, . ., K$
 \State\tab[.25cm] FOR  $j=1, . ., I$
\State\tab[.375cm]$g=g+1$
\State\tab[.5cm] FOR  $i=1, . ., r$
\State\tab[.75cm] $\mathds{B}_g \leftarrow [ \mathds{B}_g  \; \; \mathds{X}_k(j,i)  ]$   
\State\tab[.75cm] Find $\mathcal{J}_i$: set of other UEs receiving the same 
\State\tab[.65cm] interference signal $\mathds{X}_k(j,i)$, $|\mathcal{J}_i|=(L-|S|-1)$. \tab[.65cm] Sort UEs in $\mathcal{J}_i$ in ascending order.
\State \tab[.75cm]For each user in $\mathcal{J}_i$, find interference vector $\mathbf{x}_k^q$,  \tab[.60cm]s.t. 
 $\mathrm{UE}_k\in \mathcal{J}_i$ and $\mathds{X}_k(j,i)\not\in \mathbf{x}_k^q$.  
 \State \tab[.75cm] $\mathds{Q}_i \leftarrow$ set of vectors $\mathbf{x}_k^q$ 
\State\tab[.5cm] END FOR
\State \tab[1.10cm]If $|\mathcal{J}_i|\geq 1$
\State\tab[1.40cm] FOR $R=1,\ldots , |\mathcal{J}_i|$
 \State\tab[1.70cm]  FOR $e=1, \ldots, |\mathds{Q}_1(:,R)|$  $\; \;$ 
\State\tab[2.20cm]FOR $c=1, \ldots, |\mathds{Q}_2(:,R)|$
 \State\tab[2.70cm]IF  $\mathds{Q}_1(e,R) =\mathds{Q}_2(c,R)$
\State \tab[3cm]$\mathds{B}_g=[   \mathds{B}_g\; \; \mathds{Q}_1(e,R)]$
\State\tab[3cm]Go to 21, i.e., next iteration of $R$.
\State \tab[2.70cm]END IF
\State \tab[2.20cm] END FOR
 \State  \tab[1.70cm] END FOR
\State \tab[1.40cm] END FOR
\State  \tab[1.10cm] END IF
\tab[.25cm] $\mathds{C}_g \leftarrow \bigcup\limits_{k:\mathcal{\hat{S}}\in \mathds{X}_k} \mathrm{u}_k,$ for  $\hat{S}\subseteq \mathds{B}_g$, where $|\hat{S}| = r$
\State\tab[3.10cm]   FOR $e=1, \ldots, |\mathds{C}_g|$
\State  \tab[3.40cm] FOR  $i=1, \ldots, r$
\State\tab[3.40cm]$\mathds{A}_g=[\mathds{A}_g \; \;h_{\mathds{C}_g(e),\mathcal{N}_{\mathds{C}_g(e)}(i)}]$
\State\tab[3.40cm]END FOR
\State \tab[3.10cm]END FOR
\State  \tab[.25cm] Remove interference signals in $\mathds{B}_g$ from $ (\mathds{X}_k)_{k=1}^K$
\State \tab[.25cm]$\mathcal{J}_i=[\;] \; \; \mathds{Q}_i=[\;]$ $\; \;$ for $\;i= 1, \ldots, r$
\State \tab[.25cm]
\State END FOR
 \State END FOR

\end{algorithmic}
\end{algorithm}
The next step deals with the delivery from the ENs to the UEs over the partially connected interference channel. This  is the  main distinction of our work from  \cite{zewail2017coded}, where  the authors assume orthogonal links from  the relay nodes (ENs in our model) to UEs. Hence, the relay nodes simply  transmit 
 coded multicast messages $\textbf{X}_i^\mathcal{S}$  over  orthogonal links to their intended receivers. In our scheme, in order to  manage the interference between the transmitters. We use  real interference alignment  as explained in the sequel.
 
  $\textbf{Step 2}$ :  Delivery from the ENs to 
 $\mathrm{UE}_k$, $k\in[K]$, aims at delivering   the following set of  messages:
\begin{equation}
\mathcal{M}_k=\bigcup\limits_{\substack{i, \mathcal{S}: i \in \mathcal{N}_k, \mathcal{S}\subseteq[L], \\  \;|\mathcal{S}|=t_E+1, \; {\mathrm{Index}(i,k)\in \mathcal{S}} }} X_{i}^\mathcal{S},
\end{equation}
where $|\mathcal{M}_k |=r{L-1\choose t_E}$. On the other hand, the transmission of the following messages interfere with the delivery  of the messages in $\mathcal{M}_k$:
\begin{equation}
\mathcal{I}_k=\bigcup\limits_{\substack{i,  \mathcal{S}: i \in \mathcal{N}_k,  \mathcal{S}\subseteq [L], \\ \;|\mathcal{S}|=t_E+1, \; {\mathrm{Index}(i,k)\not \in  S} }} X_{i}^S.
\end{equation}

Each  $X_{i}^S\in \mathcal{I}_k$ causes interference at $L-|\mathcal{S}|$ UEs,  including  $\mathrm{UE}_k$.
 Hence, the  total number of interfering signals at   $\mathrm{UE}_k$ from the ENs in   $ \mathcal{N}_k$ is  $rI$, where $I \triangleq {L\choose t_E+1}-{L-1\choose t_E}$ is the number of interfering signals from each  EN connected to  $\mathrm{UE}_k$.

We enumerate the ENs in $\mathcal{N}_k $, $k \in [K]$, such that $\mathcal{N}_k(q)$  is  the q-th element in  $\mathcal{N}_k$  in ascending order. At  $\mathrm{UE}_k$, $k \in [K]$, we define  the $ \textit {interference matrix}$  $\mathds{X}_k$  to be an $ I \times r$  matrix whose columns are denoted by    $\{\mathbf{x}_k^q\}_{q=1}^{r}$, where the q-th column  $\mathbf{x}_k^q$  represents   the interference caused by a different EN in   $ \mathcal{N}_k(q)$. For  each column vector $\mathbf{x}_k^q$, we sort the set of interfering signals $ \mathcal{I}_k$ for $i=\mathcal{N}_k(q)$ in ascending order.  In  Example 1, we have $\mathcal{N}_1(1)=\mathrm{EN}_1$, $\mathcal{N}_1(2)=\mathrm{EN}_2$,  etc., and the interference matrices  are  shown  in  TABLE \ref{Table:3}.
\begin{table*}[]
\centering
\captionsetup{justification=raggedright,singlelinecheck=false}
\begin{adjustbox}{width=1\textwidth}
\small
\begin{tabular}{|l|l|l|l|l|l|l|l|l|l|}
\hline
\multicolumn{1}{|c|}{$\mathds{X}_1$}                                          & \multicolumn{1}{c|}{$\mathds{X}_2$}                     & \multicolumn{1}{c|}{$\mathds{X}_3$}                     & \multicolumn{1}{c|}{$\mathds{X}_4$}                     & \multicolumn{1}{c|}{$\mathds{X}_5$}                     & \multicolumn{1}{c|}{$\mathds{X}_6$}                     & \multicolumn{1}{c|}{$\mathds{X}_7$}                     & \multicolumn{1}{c|}{$\mathds{X}_8$}                     & \multicolumn{1}{c|}{$\mathds{X}_9$}                     & \multicolumn{1}{c|}{$\mathds{X}_{10}$}                    \\ \hline
\multicolumn{1}{|c|}{$\mathbf{X}_{1}^{2,3} \; \mathbf{X}_{2}^{2,3}$} & $\mathbf{X}_{1}^{1,3} \; \mathbf{X}_{3}^{2,3}$ & $\mathbf{X}_{1}^{1,2} \; \mathbf{X}_{4}^{2,3}$ & $\mathbf{X}_{1}^{1,2} \; \mathbf{X}_{5}^{2,3}$ & $\mathbf{X}_{2}^{1,3} \; \mathbf{X}_{3}^{1,3}$ & $\mathbf{X}_{2}^{1,2} \; \mathbf{X}_{4}^{1,3}$ & $\mathbf{X}_{2}^{1,2} \; \mathbf{X}_{5}^{1,3}$ & $\mathbf{X}_{3}^{1,2} \; \mathbf{X}_{4}^{1,2}$ & $\mathbf{X}_{3}^{1,2} \; \mathbf{X}_{5}^{1,4}$ & $\mathbf{X}_{4}^{1,2} \; \mathbf{X}_{5}^{1,2}$ \\
$\mathbf{X}_{1}^{2,4} \; \mathbf{X}_{2}^{2,4}$                       & $\mathbf{X}_{1}^{1,4} \; \mathbf{X}_{3}^{2,4}$ & $\mathbf{X}_{1}^{1,4} \; \mathbf{X}_{4}^{2,4}$ & $\mathbf{X}_{1}^{1,3} \; \mathbf{X}_{5}^{2,4}$ & $\mathbf{X}_{2}^{1,4} \; \mathbf{X}_{3}^{1,4}$ & $\mathbf{X}_{2}^{1,4} \; \mathbf{X}_{4}^{1,4}$ & $\mathbf{X}_{2}^{1,3} \; \mathbf{X}_{5}^{1,4}$ & $\mathbf{X}_{3}^{1,4} \; \mathbf{X}_{4}^{1,4}$ & $\mathbf{X}_{3}^{1,3} \; \mathbf{X}_{5}^{1,4}$ & $\mathbf{X}_{4}^{1,3} \; \mathbf{X}_{5}^{1,3}$ \\
$\mathbf{X}_{1}^{3,4} \; \mathbf{X}_{2}^{3,4}$                       & $\mathbf{X}_{1}^{2,4} \; \mathbf{X}_{3}^{3,4}$ & $\mathbf{X}_{1}^{3,4} \; \mathbf{X}_{4}^{3,4}$ & $\mathbf{X}_{1}^{2,3} \; \mathbf{X}_{5}^{3,4}$ & $\mathbf{X}_{2}^{3,4} \; \mathbf{X}_{3}^{3,4}$ & $\mathbf{X}_{2}^{2,4} \; \mathbf{X}_{4}^{3,4}$ & $\mathbf{X}_{2}^{2,3} \; \mathbf{X}_{5}^{3,4}$ & $\mathbf{X}_{3}^{2,4} \; \mathbf{X}_{4}^{2,4}$ & $\mathbf{X}_{3}^{2,3} \; \mathbf{X}_{5}^{2,4}$ & $\mathbf{X}_{4}^{2,3} \; \mathbf{X}_{5}^{2,3}$ \\ \hline
\end{tabular}
\caption{The interference matrices at the  UEs of  Example 1.}
\label{Table:3}
\end{adjustbox}
\end{table*}
We will use real IA,  presented in \cite{6846359} and  extended to complex channels in \cite{5513550},  for the delivery  from the ENs to the  UEs to align each of the $r$ interfering signals  in $\mathcal{I}_k$, one from each EN, in  the same subspace. We define $\mathds{A}$,  $\mathds{B}$  and  $\mathds{C}$ to be the basis matrix, i.e., function of the channel coefficients, the data matrix and user matrix, respectively, where  the dimensions of these matrices are $G\times r{r+L-|S|-1\choose r}$, $G\times (r+L-|S|-1)$ and $G\times {r+L-|S|-1\choose r}$, respectively, where $G={H \choose t_E+1}$. We denote  the rows of these matrices  by $\mathds{A}_g$,  $\mathds{B}_g$ and   $\mathds{C}_g$, respectively, where $g\in[G]$. The row vectors  $\{\mathbf{\mathds{A}}_g\}_{g=1}^G$   are used to generate the set of monomials $\mathcal{G}(\mathds{A}_g)_{g=1}^G$. Note that, the function $\mathcal{T}(u)$ defined in  \cite{7282567} corresponds to $\mathcal{G}(\mathds{A}_g)$ in our notation.  The set $\mathcal{G}(\mathds{A}_g)_{g=1}^G$ is used as the transmission directions for the modulation constellation $\mathbb{Z}_Q$ \cite{7282567} for the whole network. In other words,  each row data vector $\mathds{B}_g$  will use the set $\mathcal{G}(\mathds{A}_g)$ as the transmission directions of all its data to align all the  $r$ interfering signals from  $\mathds{B}_g$ in  the same subspace at  $\mathrm{UE}_k\in \mathds{C}_g$, if these $r$ signals   belong to  $\mathds{X}_k$. 

We next explain matrix $\mathds{C}$ more clearly. For each $\mathcal{\hat{S}}\subseteq \mathds{B}_g$ with $|\mathcal{\hat{S}}| = r$, there will be a  user at which these data will be aligned in the same subspace, i.e., $|\mathds{C}_g|={r+L-|S|-1\choose r}$. The row $\mathds{C}_g$ consists of  $\mathrm{UE}_k$,  where $ \hat{S}\in \mathds{X}_k$.

We employ  Algorithm 1 to obtain matrices $\mathds{A}$,  $\mathds{B}$  and  $\mathds{C}$ for a receiver connectivity of $r=2$, and for arbitrary receiver connectivity when $t_E= L-2$. In  Example 1, the three matrices  are given  as follows:

\begin{algorithm}[h] \footnotesize
  Generating the first rows  for $\mathds{A}$, $\mathds{B}$ and $\mathds{C}$ Matrices in Example 1
  \begin{algorithmic}[1]

 \State FOR  $k=1$ and $ j=1$ 
 \State\tab[.5cm] FOR  $i=1$ 
 \State\tab[.75cm] $\mathds{B}_1=[\mathbf{X}_{1}^{2,3} ]$   
 \State \tab[.75cm]  $\mathcal{J}_1=[\mathrm{UE}_4]$ and 
$\mathds{Q}_1=\begin{pmatrix}
\mathbf{X}_{5}^{2,3} \\
\mathbf{X}_{5}^{2,4} \\
\mathbf{X}_{5}^{3,4}  
\end{pmatrix},$
 \State\tab[.5cm] FOR  $i=2$ 
 \State\tab[.75cm] $\mathds{B}_1=[ \mathbf{X}_{1}^{2,3}   \mathbf{X}_{2}^{2,3}  ]$   
 \State\tab[.75cm]  $\mathcal{J}_2=[\mathrm{UE}_7]$ and 
$\mathds{Q}_2=\begin{pmatrix}
\mathbf{X}_{5}^{1,3}\\
\mathbf{X}_{5}^{1,4}\\
\mathbf{X}_{5}^{3,4} \\
\end{pmatrix},$
 \State\tab[1.40cm] FOR $R=1$
 \State \tab[1.50cm] The two loops in line  14 and 15 in Algorithm 1 are used to find the common message in the 2 sets $\mathds{Q}_1$ and  \tab[1.50cm] $\mathds{Q}_2$   
 \State \tab[1.50cm]  $\mathds{Q}_1(1,3) =\mathds{Q}_2(1,3)$
 \State \tab[1.50cm]$\mathds{B}_1=[ \mathds{X}_1^{2,3} \;\;  \mathds{X}_2^{2,3}    \;\; \mathds{X}_5^{3,4}   ]$ 
 \State\tab[.25cm] $\mathds{C}_1= [ \mathrm{UE}_1 \; \; \mathrm{UE}_4 \; \; \mathrm{UE}_5]$
 \State\tab[3.10cm]   FOR $e=1, \ldots, 3$
 \State  \tab[3.40cm] FOR  $i=1, \ldots, 2$
 \State\tab[3.40cm]$\mathds{A}_1=[\mathds{A}_1 \; \;h_{\mathds{C}_1(e),\mathcal{N}_{\mathds{C}_1(e)}(i)}]$
 \State\tab[3.40cm]END FOR
 \State \tab[3.10cm]END FOR
 \State \tab[.25cm] $\mathds{A}_1= [ h_{1,1} 
 \; h_{1,2} \; h_{4,1}\; h_{4,5} \; h_{7,2}\; h_{7,5}]$
 \State  \tab[.25cm] Remove interference signals in $\mathds{B}_1$ from $ \mathds{X}_1$
  \State\tab[.25cm]$\mathcal{J}_i=[\;] \; \; \mathds{Q}_i=[\;]$ $\; \;$ for $\;i= 1, 2$
  \State\tab[.25cm]END FOR
  \end{algorithmic}
\end{algorithm}
\[\footnotesize
\myMatrix{\mathds{A}} = \begin{pmatrix}
h_{1,1} &h_{1,2}  &h_{4,1}&h_{4,5} &h_{7,2}&h_{7,5}\\
h_{1,1} &h_{1,2}  &h_{3,1}&h_{3,4} &h_{6,2}&h_{6,4}\\
h_{1,1} &h_{1,2}  &h_{2,1}&h_{2,3} &h_{5,2}&h_{5,3}\\
h_{2,1} &h_{2,3}  &h_{4,1}&h_{4,5} &h_{9,3}&h_{9,5}\\
h_{2,1} &h_{2,3}  &h_{3,1}&h_{3,4} &h_{8,3}&h_{8,4}\\
h_{3,1} &h_{3,4}  &h_{4,1}&h_{4,5} &h_{10,4}&h_{10,5}\\
h_{5,2} &h_{5,3}  &h_{7,2}&h_{7,5} &h_{9,3}&h_{9,5}\\
h_{5,2} &h_{5,3}  &h_{6,2}&h_{6,4} &h_{8,3}&h_{8,4}\\
h_{6,2} &h_{6,4}  &h_{7,2}&h_{7,5} &h_{10,4}&h_{10,4}\\
h_{8,3} &h_{8,4}  &h_{10,4}&h_{10,5} &h_{9,3}&h_{9,5}\\
\end{pmatrix},
\]
\[\footnotesize
\myMatrix{\mathds{B}} = \begin{pmatrix}
\mathbf{X}_{1}^{2,3}  &\mathbf{X}_{2}^{2,3} & \mathbf{X}_{5}^{3,4} \\
\mathbf{X}_{1}^{2,4}  &\mathbf{X}_{2}^{2,4} & \mathbf{X}_{4}^{3,4} \\
\mathbf{X}_{1}^{3,4}  &\mathbf{X}_{2}^{3,4} & \mathbf{X}_{3}^{3,4} \\
\mathbf{X}_{1}^{1,3}  &\mathbf{X}_{3}^{2,3} & \mathbf{X}_{5}^{2,4} \\
\mathbf{X}_{1}^{1,4}  &\mathbf{X}_{3}^{2,4} & \mathbf{X}_{4}^{2,4} \\
\mathbf{X}_{1}^{1,2}  &\mathbf{X}_{4}^{2,3} & \mathbf{X}_{5}^{2,3} \\
\mathbf{X}_{2}^{1,3}  &\mathbf{X}_{3}^{1,3} & \mathbf{X}_{5}^{1,4} \\
\mathbf{X}_{2}^{1,4}  &\mathbf{X}_{3}^{1,4} & \mathbf{X}_{4}^{1,4} \\
\mathbf{X}_{2}^{1,2}  &\mathbf{X}_{4}^{1,3} & \mathbf{X}_{5}^{1,3} \\
\mathbf{X}_{4}^{1,2}  &\mathbf{X}_{3}^{1,2} & \mathbf{X}_{5}^{1,2} \\
\end{pmatrix}
, \;
\myMatrix{\mathds{C}} = \begin{pmatrix}
\mathrm{UE}_1  &\mathrm{UE}_4 & \mathrm{UE}_7 \\
\mathrm{UE}_1  &\mathrm{UE}_3 & \mathrm{UE}_6  \\
\mathrm{UE}_1  &\mathrm{UE}_2 & \mathrm{UE}_5  \\
\mathrm{UE}_2  &\mathrm{UE}_4 & \mathrm{UE}_9  \\
\mathrm{UE}_2  &\mathrm{UE}_3 & \mathrm{UE}_8  \\
\mathrm{UE}_3  &\mathrm{UE}_4 & \mathrm{UE}_{10}  \\
\mathrm{UE}_5  &\mathrm{UE}_7 & \mathrm{UE}_9  \\
\mathrm{UE}_5  &\mathrm{UE}_6 & \mathrm{UE}_8  \\
\mathrm{UE}_6  &\mathrm{UE}_7 & \mathrm{UE}_{10}  \\
\mathrm{UE}_8  &\mathrm{UE}_{10} & \mathrm{UE}_9 \\
\end{pmatrix}.
\]

 Then, for each signal in   $\mathds{B}_g$, we construct a constellation that is scaled by the monomial set $\mathcal{G}(\mathds{A}_g)$, i.e, the signals   $\mathbf{X}_{2}^{2,4}$ in $\mathds{B}_2$  use the monomials in    $\mathcal{G}(\mathds{A}_2)$, resulting in the  signal constellation
$\sum_{v\in \mathcal{G} (\mathds{A}_g)} v\mathbb{Z}_Q.$

Focusing on  Example 1, we want to assess whether the
interfering signals have been aligned, and whether  the requested
subfiles arrive with independent channel coefficients, 
the decodability is guaranteed. Starting with $\mathrm{u}_1$, the received constellation corresponding to  the desired signals  $\mathbf{X}_{1}^{1,2}$, $\mathbf{X}_{1}^{1,3}$,  $\mathbf{X}_{1}^{1,4}$, $\mathbf{X}_{2}^{1,2}$, $\mathbf{X}_{2}^{1,3}$ and $\mathbf{X}_{2}^{1,4}$:
\begin{equation}
\label{4}
\begin{split}
 C_D&=h_{1,1}\sum_{v\in \mathcal{G} (\mathds{A}_6)} v\mathbb{Z}_Q +h_{1,1}  \sum_{v\in \mathcal{G} (\mathds{A}_4)} v\mathbb{Z}_Q+ h_{1,1} \sum_{v\in \mathcal{G} (\mathds{A}_5)} v\mathbb{Z}_Q \\ &+h_{1,2}  \sum_{v\in \mathcal{G} (\mathds{A}_9)} v\mathbb{Z}_Q +h_{1,2}  \sum_{v\in \mathcal{G} (\mathds{A}_7)} v\mathbb{Z}_Q + h_{1,2} \sum_{v\in \mathcal{G} (\mathds{A}_8)} v\mathbb{Z}_Q.
\end{split}
\end{equation}
The received constellation for the interfering signals  $\mathbf{X}_{1}^{2,3}$ $\mathbf{X}_{2}^{2,3}$,  $\mathbf{X}_{1}^{2,4}$, $\mathbf{X}_{2}^{2,4}$, $\mathbf{X}_{2}^{3,4}$ and $\mathbf{X}_{2}^{3,4}$ is 
\begin{equation}
\label{5}
\begin{split}
 C_I&=h_{1,1}\sum_{v\in \mathcal{G} (\mathds{A}_1)} v\mathbb{Z}_Q +h_{1,2} \sum_{v\in \mathcal{G} (\mathds{A}_1)} v\mathbb{Z}_Q+ h_{1,1}\sum_{v\in \mathcal{G} (\mathds{A}_2)} v\mathbb{Z}_Q \\ &+h_{1,2}  \sum_{v\in \mathcal{G} (\mathds{A}_2)} v\mathbb{Z}_Q + h_{1,1} \sum_{v\in \mathcal{G} (\mathds{A}_3)} v\mathbb{Z}_Q +h_{1,2}  \sum_{v\in \mathcal{G} (\mathds{A}_3)} v\mathbb{Z}_Q.
\end{split}
\end{equation}



\begin{figure*}
\centering
\captionsetup{justification=raggedright,singlelinecheck=false}
\includegraphics[width=27cm,height=11cm,keepaspectratio]{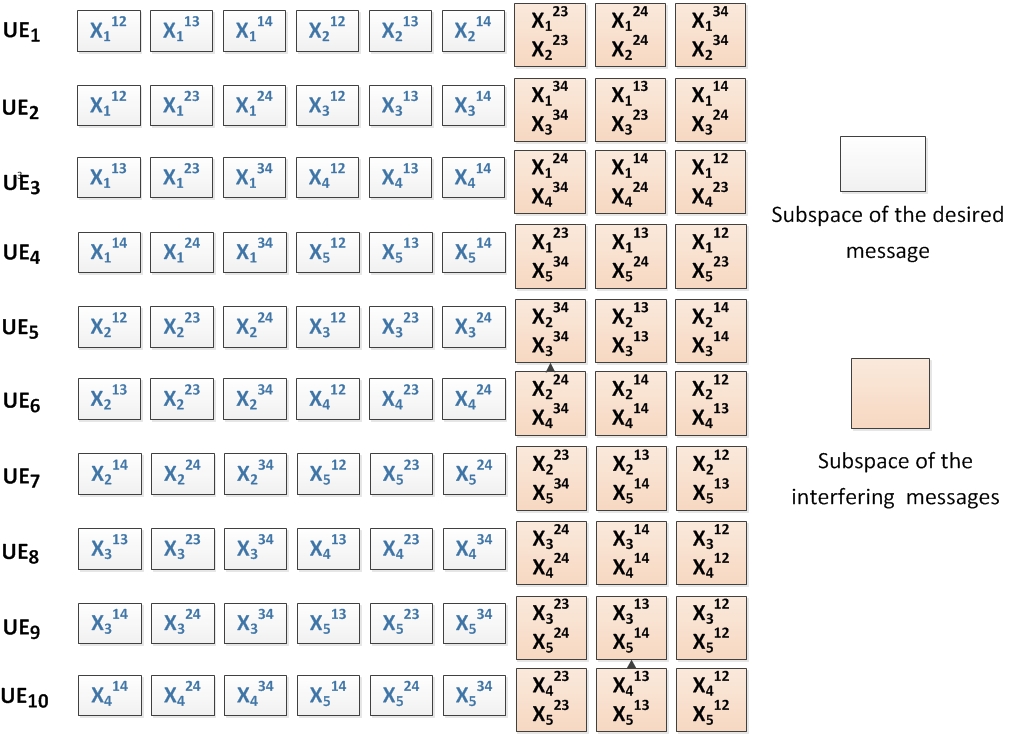}
\caption{  The signal space  for  UEs in Example 1 after applying  real IA. }
\label{IA Signal space}
\end{figure*}

Eqn.  \eqref{5} proves that every two interfering signals, one from each EN, i.e., the first two terms in Eqn. \eqref{5},  have collapsed into the same sub-space. Also, since the monomials $\mathcal{G} (\mathds{A}_1)$, $\mathcal{G} (\mathds{A}_2)$ and $\mathcal{G} (\mathds{A}_3)$ do not  overlap and linear independence is obtained, the interfering signals will align in $I=3$ different sub-spaces. 
 We can  also see   in \eqref{4}  that the  monomials corresponding to the intended messages  do not align, and  rational  independence is guaranteed (with high probability), and the desired signals will be received over 6 different subspaces. Since the monomials form different constellations, $C_D$ and $C_I$, whose terms are functions of different channel coefficients,
we can assert that these monomials do not overlap. Hence, we can claim that  IA is achieved. 
Our scheme guarantees that  the desired signals at each user will be received in $r {L-1\choose t_E}$ different subspaces, and each $r$ interfering  signals will be aligned into the same subspace, i.e., one from each EN, resulting in  a total of $I= {L\choose t_E+1}- {L-1\choose t_E}$ interference subspaces. The signal space  for  UEs in Example 1 after applying  real IA   is given in Fig. \ref{IA Signal space}.

When $t_E=L-1$, the number of interference signals at each user is $I=0$. Hence, we just transmit the constellation points corresponding to each signal. We are sure that the decodability is guaranteed since all channel coefficients are i.i.d. according to a continuous  distribution. 

 $\mathrm{UE}_k$ utilizes  its cache content  $Z_k$ to extract   $f^i_{k, \mathcal{T}}$, for  $i\in  \mathcal{N}_k$ and $\mathrm{Index}(i,k)\not \in  \mathcal{T}$. Therefore,  $\mathrm{UE}_k$ reconstructs $f^i_{k}$ and
decodes its requested file $W_{k}$. In Example 1, $\mathrm{UE}_1$ utilizes its memory $Z_1$ in  TABLE \ref{Table:a} to extract $f^i_{1, \mathcal{T}}$, for $i=1,2$, and $\mathcal{T}=\{2, 3, 4\}$. Hence,  $\mathrm{UE}_1$  reconstructs $f^1_{1}$ and $f^2_{1}$,  and decodes its requested file   $W_{1}$; and similarly for the remaining UEs. 
Thus, the edge NDT  from ENs to the   UEs   is equal to  $\delta_E (\frac{1}{4}, 0, \rho) = \frac{9}{8}$, while the total NDT  is $\delta (\frac{1}{4}, 0, \rho)=\frac{3}{4\rho}+\frac {9}{8}$. In the general case, 
the  NDT from the  ENs to the UEs by using the MDS-IA scheme  is given by
\begin{equation}
 \delta_E(\mu_R, 0, \rho) =\frac{{L-1\choose t_E}(r-1) + {L\choose t_E+1} }{r{L\choose t_E} }=\frac{L-t_E}{r}\left(\frac{r-1}{L} + \frac{1}{t_E+1} \right) .
\end{equation}
Together with the fronthaul NDT in (\ref{Fron}), we obtain the end-to-end NDT in Theorem 1.  NDT achieved by the MDS-IA  for various system parameters is  presented in Section \ref{Numerical Results}.
\subsection{MDS-IA Scheme with  Cache Enabled  ENs}
IN  the  MDS-IA scheme,  when $ \mu_T\geq \frac{1}{r}$, i.e., each $ \mathrm{EN}_i$, $i\in [K]$, can cache the encoded subfile $f_n^i$, $\forall n$, and  the users' requests  can be satisfied without the participation of the cloud server in  the delivery phase,  i.e., $\rho=0$. In this case  each $\mathrm{EN}_i$, $i\in [K]$,  can act as  a server for its connected UEs,  and we  can use the same placement  scheme in Section \ref{Scheme1}-\ref{MDS-IA no cache} for the UEs,   and the same delivery  scheme  in  Section \ref{Scheme1}-\ref{MDS-IA no cache}- Step 2,  given that the coded multicast messages in \eqref{coded-multicasting}   can now  be generated locally at the ENs. Then,  end-to-end achievable NDT  is  given by 
\begin{equation}
\label{N}
 \delta_{\text{MDS-IA}} (\mu_R, \mu_T, \rho)= \frac{L-t_E}{r}\left[ \frac{r-1}{L} +  \frac{1}{t_E+1} \right]
 \end{equation}.
\subsubsection{ \textbf{Cache Placement Phase}}
In the following, we extend the  proposed
scheme to the case with  $  0 \leq \mu_T <1/r$. We use a similar  cache placement  scheme to the one in [21]. We form the coded chunks  $f_n^i$ as in Section  \ref{Scheme1}-\ref{MDS-IA no cache}.  Then, the server divides each chunk  into two parts, $f_n^{i,1}$ and $f_n^{i,2}$ ,  with sizes  $\mu_T F$   and $( \frac{1}{r}-\mu_T) F$, respectively.  The  cloud server places $f_n^{i,1}$, $\forall n $,  in the cache memory of $\mathrm{EN}_i$, where the total size of the cached pieces is $\mu_TNF $ , satisfying the  cache memory constraint with equality. 
  The  UE cache  placement scheme is the same as  in Section \ref{Scheme1}-\ref{MDS-IA no cache}.      At the end of the placement phase, each user stores $Nr\binom {L-1} {t_E-1}$ pieces from each  set of  the encoded chunks, which sum up to $\mu_RNF $, satisfying the UE cache  memory constraint with equality. 
\subsubsection{ \textbf{Delivery Phase}}
Delivery phase is divided into two parts. In the first part,  the cloud server   delivers the subfiles in 
 \begin{equation}
 \label{w}
 \{f^{i,2}_{d_k,\mathcal{T}} : k \notin \mathcal{T}, \mathcal{T} \subseteq[L], |\mathcal{T}|= t_E\}
 \end{equation}
to  $\mathrm{UE}_k$, i.e., the subfiles  of  $ f^{i,2}_{d_k}$ that have not been already stored in  the cache memories of   $\mathrm{UE}_k$ and $\mathrm{EN}_i$. The total number of such subfiles is  ${L\choose t_E}- \binom{L-1}{t_E-1}$. For these,  we use the same delivery scheme  in Section \ref{Scheme1}-\ref{MDS-IA no cache}-2. The achievable NDT is given by
\begin{equation}
\label{ax-M_T}
 \delta_{\text{MDS-IA}}(\mu_R, \mu_T, \rho)  =  \left(\frac{1}{r}-\mu_T \right)  \left(L-t_E\right) \left[ \frac{r-1}{L} +  \frac{1}{t_E+1}\left(1+\frac{1}{\rho} \right) \right],
 \end{equation}
 where the factor $(\frac{1}{r}-\mu_T ) $ is   due to the reduction in    the size of  the coded multicast message from the cloud server  to the ENs thanks to the already cached contents in EN caches.
 
In the second part,  ENs  deliver the subfiles in 
 \begin{equation}
 \label{w}
 \{f^{i,1}_{d_k,\mathcal{T}} : k \notin \mathcal{T}, \mathcal{T} \subseteq[L], |\mathcal{T}|= t_E\}
 \end{equation}
to  $\mathrm{UE}_k$, i.e., the subfiles  of file $ f^{i,1}_{d_k}$ which have been  cached by  $\mathrm{EN}_i$ but not by$\mathrm{UE}_k$. The total number of such subfiles is  ${L\choose t_E}- \binom{L-1}{t_E-1}$. For this case,  we use the  delivery scheme from the ENs to the UEs  presented  in Section \ref{Scheme1}-\ref{MDS-IA no cache}-2, where  the coded multicaste message 
 \begin{equation}
\textbf{X}_{i,1}^\mathcal{S} \triangleq \bigoplus_{k: k\in \mathcal{K}_i, Index(i, k)\in \mathcal{S}} f^{i,1}_{{d_k}, \mathcal{S} \backslash \mathrm{Index}(i, k)}.
\end{equation}
will be generated locally at  $\mathrm{EN}_i$, where $|\textbf{X}_{i,1}^\mathcal{S}|=  \frac{ \mu_T  F}{{L\choose t_E}}  $ bits. Accordingly,  the fronthaul NDT from the cloud server to the ENs is zero.   By following the same approach of  Section \ref{Scheme1}-\ref{MDS-IA no cache} Step 2, the achievable NDT from the ENs to the UEs is given by
 \begin{equation}
 \delta_E(\mu_R, \mu_T, \rho) = \mu_T \frac{{L-1\choose t_E}(r-1) + {L\choose t_E+1} }{{L\choose t_E} }= \mu_T  \left(L-t_E\right)\left(\frac{r-1}{L} + \frac{1}{t_E+1} \right) .
\end{equation}
Together with the  NDT in \eqref{ax-M_T}, we obtain the end-to-end NDT as given in Theorem 1.

\section{Soft-Transfer Scheme}\label{Scheme}

  As in the MDS-IA scheme it is easier and more intuitive to first introduce  a  RACN  architecture without cache memories at the ENs. Generalization to cache-enabled ENs will follow easily from this initial scheme.
\subsection{Soft-Transfer Scheme with out Cache Memories at the ENs}\label{subScheme}
Here, we present a centralized  caching
scheme for     with receiver connectivity $r$, and   $t_U\in[K]$  when there is no cache memory at the ENs.  The soft-transfer of channel input symbols over fronthaul links is proposed in \cite{ Simeone2009}, where the cloud  server implements ZF-beamforming and quantizes the encoded signal to be transmitted to each EN. Therefore, the fronthaul NDT is given by
\begin{equation}
\label{Soft}
\delta_{F-\text{soft}}(\mu_R, \mu_T, \rho) =\left(1-\frac{t_U}{K} \right) \frac{K}{H \rho},
\end{equation} 
while the total NDT can be expressed  as 
\begin{equation}
\label{TotalSoft}
\delta_{\text{soft}}(\mu_R, \mu_T, \rho) =\delta_{E-\text{Ideal}} + \delta_{F-\text{soft}},
\end{equation}
where $\delta_{E-\text{Ideal}}$ is  the achievable edge NDT in an ideal system in which the ENs can acts as one big multi-antenna  transmitter. This is equivalent to assuming   that the whole library $\mathcal{W}$ can be cached  at all the ENs; and hence,  full cooperation 
 among the ENs is possible for any user demand vector.   We will provide a coding scheme that uses ZF  for this  ideal system to provide a general expression for  $\delta_{E-\text{Ideal}}$.

\subsubsection{\textbf{Cache Placement Phase}}\label{Soft placement}

For any file $W_n$ in the library, $n \in [N]$, we partition it into
 ${K\choose t_U}$ equal-size subfiles, each of  which is denoted by   $W_{n, \mathcal{T}}$, where  $\mathcal{T}\subseteq[K]$, $|\mathcal{T}|=t_U$.  The subfiles   $W_{n, \mathcal{T}}$, $\forall n$, are stored in the cache memory of  $\mathrm{UE}_k$ if $ k\in \mathcal{T}$. 
 At the end of the  placement phase, each user stores
$N\binom {K-1} {t_U-1}$ subfiles, each of size $\frac{F}{{K\choose t_U}}$
bits, which sum up to $MF$ bits,  satisfying the cache capacity constraint with equality.

\textbf{Example 2.}
 $\;$ Consider the   $4\times 6$  RACN  architecture with $H=4$, $K=N=6$,  $r=2$, $\mu_T=0$ bits and $L=3$. For $t_U=2$, file $W_n$, $\forall n \in [N]$, is divided  into ${6\choose 2}=15$  disjoint subfiles  $W_{n, \mathcal{T}}$, where $\mathcal{T}\subseteq[K]$ and $|\mathcal{T}|= t_U =2$. The size of each subfile is $\frac{F}{15}$ bits. Cache contents of each user are listed in TABLE \ref{Table:qqq}. Observe that each user stores $6\binom {5} {1}=30$ subfiles, each of size $\frac{F}{15}$ bits, which sum up to $2F$ bits,  satisfying the memory constraint with equality.

\begin{table}[]
\begin{adjustbox}{width=.65\textwidth}
\small
\begin{tabular}{|c|l|l|l|l|l|l|}
\hline
\multicolumn{1}{|l|}{User} & \multicolumn{1}{c|}{$\mathrm{UE}_{1}$} & \multicolumn{1}{c|}{$\mathrm{UE}_{2}$} & \multicolumn{1}{c|}{$\mathrm{UE}_{3}$} & \multicolumn{1}{c|}{$\mathrm{UE}_{4}$} & \multicolumn{1}{c|}{$\mathrm{UE}_{5}$} &\multicolumn{1}{c|}{ $\mathrm{UE}_{6}$ }\\ \hline
\multirow{5}{*}{Cache}     & \multicolumn{1}{c|}{$W_{n,12}$}       & \multicolumn{1}{c|}{$W_{n,12}$}        &\multicolumn{1}{c|}{$W_{n,13}$}                         &\multicolumn{1}{c|}{$W_{n,14}$}             &\multicolumn{1}{c|}{$W_{n,15}$}       &\multicolumn{1}{c|}{$W_{n,16}$}         \\ \cline{2-7} 
                           & \multicolumn{1}{c|}{$W_{n,13}$}       & \multicolumn{1}{c|}{$W_{n,23}$}                               & \multicolumn{1}{c|}{$W_{n,23}$}         & \multicolumn{1}{c|}{$W_{n,24}$}        &\multicolumn{1}{c|}{$W_{n,25}$}       & \multicolumn{1}{c|}{$W_{n,26}$}        \\ \cline{2-7} 
                           & \multicolumn{1}{c|}{$W_{n,14}$}       & \multicolumn{1}{c|}{$W_{n,24}$}                               &\multicolumn{1}{c|}{$W_{n,34}$}        &\multicolumn{1}{c|}{$W_{n,34}$}        &\multicolumn{1}{c|}{$W_{n,35}$}      & \multicolumn{1}{c|}{$W_{n,36}$}      \\ \cline{2-7} 
                           & \multicolumn{1}{c|}{$W_{n,15}$}         & \multicolumn{1}{c|}{$W_{n,25}$}                              & \multicolumn{1}{c|}{$W_{n,35}$}           & \multicolumn{1}{c|}{$W_{n,45}$}       & \multicolumn{1}{c|}{$W_{n,45}$}      & \multicolumn{1}{c|}{$W_{n,46}$}       \\ \cline{2-7} 
                           &\multicolumn{1}{c|}{$W_{n,16}$}          & \multicolumn{1}{c|}{$W_{n,26}$}                         & \multicolumn{1}{c|}{$W_{n,36}$}           & \multicolumn{1}{c|}{$W_{n,46}$}        &\multicolumn{1}{c|}{$W_{n,56}$}       &\multicolumn{1}{c|}{$W_{n,56}$}      \\ \hline
\end{tabular}

\caption{Cache contents after the placement phase for the RACN  scenario considered in Example 2, where $K=N=6$, $r=2$, $L=3$,  $t_U=2$ and $\mu_R=\frac{1}{3}$. }
  \label{Table:qqq}
 \end{adjustbox}
\end{table}
 
 \begin{table*}[]
\centering
\captionsetup{justification=raggedright,singlelinecheck=false}
\begin{adjustbox}{width=1\textwidth}
\small
\begin{tabular}{|c|l|l|l|l|l|l|l|l|l|l|l|l|}
\hline
\multicolumn{1}{|c|}{User}                                                     & \multicolumn{2}{c|}{$\mathrm{UE}_{1}$} & \multicolumn{2}{c|}{$\mathrm{UE}_{2}$} & \multicolumn{2}{c|}{$\mathrm{UE}_{3}$} & \multicolumn{2}{c|}{$\mathrm{UE}_{4}$} & \multicolumn{2}{c|}{$\mathrm{UE}_{5}$} & \multicolumn{2}{c|}{$\mathrm{UE}_{6}$} \\ \hline
\multirow{5}{*}{\begin{tabular}[c]{@{}c@{}}	Missing\\  subfiles\end{tabular}} &\multicolumn{1}{c|}{ $W_{1,23}$   }     & \multicolumn{1}{c|}{ $W_{1,35}$}        &  \multicolumn{1}{c|}{$W_{2,13}$ }       &  \multicolumn{1}{c|}{$W_{2,35}$ }       &  \multicolumn{1}{c|}{$W_{3,12} $}       &  \multicolumn{1}{c|}{$W_{3,25}$  }     & \multicolumn{1}{c|}{ $W_{4,12}$ }       & \multicolumn{1}{c|}{ $W_{4,25}$ }       & \multicolumn{1}{c|}{ $W_{5,12}$  }      &  \multicolumn{1}{c|}{$W_{5,24}$ }       & \multicolumn{1}{c|}{ $W_{6,12}$}        &  \multicolumn{1}{c|}{$W_{6,24}$}        \\ \cline{2-13} 
                                                                               & \multicolumn{1}{c|}{$W_{1,24}$  }      &  \multicolumn{1}{c|}{$W_{1,36}$ }       & \multicolumn{1}{c|}{ $W_{2,14}$ }       &  \multicolumn{1}{c|}{$W_{2,36}$   }     &  \multicolumn{1}{c|}{$W_{3,14} $ }      &  \multicolumn{1}{c|}{$W_{3,26}$  }
     & \multicolumn{1}{c|}{ $W_{4,13}$ }       & \multicolumn{1}{c|}{ $W_{4,26}$  }      & \multicolumn{1}{c|}{ $W_{5,13}$   }     & \multicolumn{1}{c|}{ $W_{5,26}$  }      & \multicolumn{1}{c|}{ $W_{6,13}$ }       & \multicolumn{1}{c|}{ $W_{6,25}$ }       \\ \cline{2-13} 
                                                                               & \multicolumn{1}{c|}{$W_{1,25}$  }      &  \multicolumn{1}{c|}{$W_{1,45}$ }       & \multicolumn{1}{c|}{ $W_{2,15}$ }       &  \multicolumn{1}{c|}{$W_{2,45}$ }       &  \multicolumn{1}{c|}{$W_{3,15} $ }     & \multicolumn{1}{c|}{ $W_{3,45}$  }     &  \multicolumn{1}{c|}{$W_{4,15}$   }     &  \multicolumn{1}{c|}{$W_{4,35}$    }    & \multicolumn{1}{c|}{ $W_{5,14}$     }   &  \multicolumn{1}{c|}{$W_{5,34}$    }    &  \multicolumn{1}{c|}{$W_{6,14}$   }     &  \multicolumn{1}{c|}{$W_{6,34}$   }     \\ \cline{2-13} 
                                                                               & \multicolumn{1}{c|}{$W_{1,26}$  }      &  \multicolumn{1}{c|}{$W_{1,46}$ }       & \multicolumn{1}{c|}{ $W_{2,16}$}        &  \multicolumn{1}{c|}{$W_{2,46}$ }       &  \multicolumn{1}{c|}{$W_{3,16}$   }      & \multicolumn{1}{c|}{ $W_{3,46}$  }     & \multicolumn{1}{c|}{ $W_{4,16}$  }      & \multicolumn{1}{c|}{ $W_{4,36}$   }     & \multicolumn{1}{c|}{ $W_{5,16}$    }    & \multicolumn{1}{c|}{ $W_{5,36}$   }     & \multicolumn{1}{c|}{ $W_{6,15}$  }      & \multicolumn{1}{c|}{ $W_{6,35}$  }      \\ \cline{2-13} 
                                                                               & \multicolumn{1}{c|}{$W_{1,34}$  }      &  \multicolumn{1}{c|}{$W_{1,56}$ }       &  \multicolumn{1}{c|}{$W_{2,34}$  }      &  \multicolumn{1}{c|}{$W_{2,56}$ }       &  \multicolumn{1}{c|}{$W_{3,24}$    }     & \multicolumn{1}{c|}{ $W_{3,56}$   }    & \multicolumn{1}{c|}{ $W_{4,23}$   }     & \multicolumn{1}{c|}{ $W_{4,56}$    }    & \multicolumn{1}{c|}{ $W_{5,23}$     }   & \multicolumn{1}{c|}{ $W_{5,46}$    }    & \multicolumn{1}{c|}{ $W_{6,23}$   }     & \multicolumn{1}{c|}{ $W_{6,45}$   }     \\ \hline
\end{tabular}
\caption{Missing  subfiles for each user's request in Example 2. These subfiles must be delivered to the corresponding user within the delivery phase.}
  \label{Table:c}
  \end{adjustbox}
\end{table*}

\subsubsection{\textbf{Delivery Phase}}\label{Soft delivery}

 Let $W_{d_k}$ denote the request of user $\mathrm{UE}_k$, $k\in[K]$. Then, the ENs need to deliver the subfiles in 
 \begin{equation}
 \label{w}
 \{W_{d_k,\mathcal{T}} : k \notin \mathcal{T}, \mathcal{T} \subseteq[K], |\mathcal{T}|= t_U\}
 \end{equation}
to  $\mathrm{UE}_k$, i.e., the subfiles  of file $W_{d_k}$ that have not been  stored in the cache of $\mathrm{UE}_k$. The total number of such subfiles is  ${K\choose t_U}- \binom{K-1}{t_U-1}$.

 \begin{table*}[ht]
\centering
\captionsetup{justification=raggedright,singlelinecheck=false}
\begin{adjustbox}{width=1\textwidth}
\small
\begin{tabular}{|c|l|l|l|l|l|l|l|l|l|l|l|l|}
\hline
\multicolumn{1}{|l|}{User}                                                     & \multicolumn{2}{c|}{$\mathrm{UE}_{1}$} & \multicolumn{2}{c|}{$\mathrm{UE}_{2}$} & \multicolumn{2}{c|}{$\mathrm{UE}_{3}$} & \multicolumn{2}{c|}{$\mathrm{UE}_{4}$} & \multicolumn{2}{c|}{$\mathrm{UE}_{5}$} & \multicolumn{2}{c|}{$\mathrm{UE}_{6}$} \\ \hline
\multirow{5}{*}{\begin{tabular}[c]{@{}c@{}}Missing \\  subfiles\end{tabular}} &\multicolumn{1}{c|}{ $W_{1,23,456}$   }     & \multicolumn{1}{c|}{ $W_{1,35,246}$}        &  \multicolumn{1}{c|}{$W_{2,13,456}$ }       &  \multicolumn{1}{c|}{$W_{2,35,146}$ }       &  \multicolumn{1}{c|}{$W_{3,12,456} $}       &  \multicolumn{1}{c|}{$W_{3,25,146}$  }     & \multicolumn{1}{c|}{ $W_{4,12,356}$ }       & \multicolumn{1}{c|}{ $W_{4,25,136}$ }       & \multicolumn{1}{c|}{ $W_{5,12,346}$  }      &  \multicolumn{1}{c|}{$W_{5,24,136}$ }       & \multicolumn{1}{c|}{ $W_{6,12,345}$}        &  \multicolumn{1}{c|}{$W_{6,24,135}$}        \\ \cline{2-13} 
                                                                               & \multicolumn{1}{c|}{$W_{1,24,356}$  }      &  \multicolumn{1}{c|}{$W_{1,36,245}$ }       & \multicolumn{1}{c|}{ $W_{2,14,356}$ }       &  \multicolumn{1}{c|}{$W_{2,36,145}$   }     &  \multicolumn{1}{c|}{$W_{3,14,256} $ }      &  \multicolumn{1}{c|}{$W_{3,26,145}$  }
     & \multicolumn{1}{c|}{ $W_{4,13,256}$ }       & \multicolumn{1}{c|}{ $W_{4,26,135}$  }      & \multicolumn{1}{c|}{ $W_{5,13,246}$   }     & \multicolumn{1}{c|}{ $W_{5,26,134}$  }      & \multicolumn{1}{c|}{ $W_{6,13,245}$ }       & \multicolumn{1}{c|}{ $W_{6,25,134}$ }       \\ \cline{2-13} 
                                                                               & \multicolumn{1}{c|}{$W_{1,25,346}$  }      &  \multicolumn{1}{c|}{$W_{1,45,236}$ }       & \multicolumn{1}{c|}{ $W_{2,15,346}$ }       &  \multicolumn{1}{c|}{$W_{2,45,136}$ }       &  \multicolumn{1}{c|}{$W_{3,15,246} $ }     & \multicolumn{1}{c|}{ $W_{3,45,126}$  }     &  \multicolumn{1}{c|}{$W_{4,15,236}$   }     &  \multicolumn{1}{c|}{$W_{4,35,126}$    }    & \multicolumn{1}{c|}{ $W_{5,14,236}$     }   &  \multicolumn{1}{c|}{$W_{5,34,126}$    }    &  \multicolumn{1}{c|}{$W_{6,14,235}$   }     &  \multicolumn{1}{c|}{$W_{6,34,125}$   }     \\ \cline{2-13} 
                                                                               & \multicolumn{1}{c|}{$W_{1,26,345}$  }      &  \multicolumn{1}{c|}{$W_{1,46,235}$ }       & \multicolumn{1}{c|}{ $W_{2,16,345}$}        &  \multicolumn{1}{c|}{$W_{2,46,135}$ }       &  \multicolumn{1}{c|}{$W_{3,16,245}$   }      & \multicolumn{1}{c|}{ $W_{3,46,125}$  }     & \multicolumn{1}{c|}{ $W_{4,16,235}$  }      & \multicolumn{1}{c|}{ $W_{4,36,125}$   }     & \multicolumn{1}{c|}{ $W_{5,16,234}$    }    & \multicolumn{1}{c|}{ $W_{5,36,124}$   }     & \multicolumn{1}{c|}{ $W_{6,15,234}$  }      & \multicolumn{1}{c|}{ $W_{6,35,124}$  }      \\ \cline{2-13} 
                                                                               & \multicolumn{1}{c|}{$W_{1,34,256}$  }      &  \multicolumn{1}{c|}{$W_{1,56,234}$ }       &  \multicolumn{1}{c|}{$W_{2,34,156}$  }      &  \multicolumn{1}{c|}{$W_{2,56,134}$ }       &  \multicolumn{1}{c|}{$W_{3,24,156}$    }     & \multicolumn{1}{c|}{ $W_{3,56,124}$   }    & \multicolumn{1}{c|}{ $W_{4,23,156}$   }     & \multicolumn{1}{c|}{ $W_{4,56,123}$    }    & \multicolumn{1}{c|}{ $W_{5,23,146}$     }   & \multicolumn{1}{c|}{ $W_{5,46,123}$    }    & \multicolumn{1}{c|}{ $W_{6,23,145}$   }     & \multicolumn{1}{c|}{ $W_{6,45,123}$   }     \\ \hline
\end{tabular}
\end{adjustbox}
\caption{The  alternative representation of the missing subfiles  of  each user in Example 2, including the UEs at which  each subfile will be zero-forced at. }
 \label{Table:d}
\end{table*}

For Example 2,  assuming, without loss of generality, that $\mathrm{UE}_k$ requests $W_k$, the missing subfiles of  each user request, to be delivered in the delivery phase, are listed in TABLE \ref{Table:c}. 
 
We first describe the delivery phase when  $ K-H \leq t_U\leq K-1$. We will later consider the  case  $t_U<K-H$ separately. Note that,  when $t_U=K$ the achievable NDT is equal to zero since each user can  cache all the $N$ files.

Case 1 (  $ K-H \leq t_U\leq K-1$ ):  We introduce an alternative representation for  each subfile in \eqref{w}, which will denote the UEs at which each of these subfiles will be zero-forced. In particular, we will denote subfile $W_{d_k,\mathcal{T}}$ by $W_{d_k,\mathcal{T},\pi }$, where $ {\pi } \subseteq[K] \backslash (\{k\}\cup\mathcal{T})$,
 $|\pi |=K-(t_U+1)$, denotes  the set of receivers at which  this subfile will be zero-forced. 
  The total number of subfiles intended for  $\mathrm{UE}_k$ is ${K\choose t_U}- \binom{K-1}{t_U-1}$ subfiles. 
TABLE \ref{Table:d} shows this alternative  representation for the  missing  subfiles for each user in Example 2.

All the  ENs   will transmit $W_{d_k,\mathcal{T} , {\pi }}$ by using the beamforming vector $\mathbf{v}_{\pi }\in \mathbb{R}^{H}$ to  zero-force  this subfile at  the   UEs in $\pi$.  We define the matrix $\mathbf{H}_{\pi }$ with dimensions $K-(t_U+1) \times H$ to be the channel matrix  from the ENs to  the UEs in  $\pi $ and the set of ENs $\mathcal{E}$.  The beamforming vector $\mathbf{v}_{\pi }$ is designed as follows:
\begin{equation}
 \mathbf{v}_{\pi }=\sum_{i=1}^{D} \mathbf{v}_i,
 \end{equation}
 where $D=H-K+(t_U+1)$ is the size of the null space of the matrix $\mathbf{H}_{\pi }$,  while $\mathbf{v}_i$ is the $i$th basis vector  of  this null space. The null space of matrix $\mathbf{H}_\pi $ with $1 \leq |\pi |\leq  H-1$  always has a non-zero element since $D\geq 1$. Hence, the subfile  $W_{d_k,\mathcal{T} , {\pi }}$  for any $\pi \geq 1$ can be always zero-forced  at the  UEs in  $\pi$. In Example 2, the size of the null space is $D=1$.

 In each  step of the delivery phase, we  transmit one subfile from each requested file, which means that we will have ${K\choose t_U}- \binom{K-1}{t_U-1}$ steps in total. The transmitted  set of subfiles  at each step will be decoded by their intended receivers   without any interference since each subfile $W_{d_k,\mathcal{T},\pi }$  in this set   is already cached at $|\mathcal{T}|=t_U$ UEs, $\mathcal{T} \subseteq \mathcal{U}$, and will be zero-forced at $|\pi |=K-(t_U+1)$ other UEs,  $\pi  \subseteq \mathcal{U}$. Since $ \mathcal{T}\cap \pi   =\phi$ and $|\mathcal{T}|+|\pi |=|\mathcal{U}|-1$,  each subfile will not cause any interference at the $|K|-1$ undesired UEs. 
 Hence, the edge NDT of the ideal system is given  by
 \begin{equation}
 \delta_{E-\text{Ideal}}(\mu_R, \mu_T, \rho)  =\frac{K- t_U}{K} .
 \end{equation}
The  end-to-end achievable  NDT  for  the soft-transfer scheme is given by 
\begin{equation}
 \delta_{\text{soft}}(\mu_R, \mu_T, \rho) =(K-t_U) \left(\frac{1}{ K}+\frac{1}{ H \rho} \right).
 \end{equation}
 
 \begin{figure*}
\centering
\captionsetup{justification=raggedright,singlelinecheck=false}
\includegraphics[width=18cm,height=8cm,keepaspectratio]{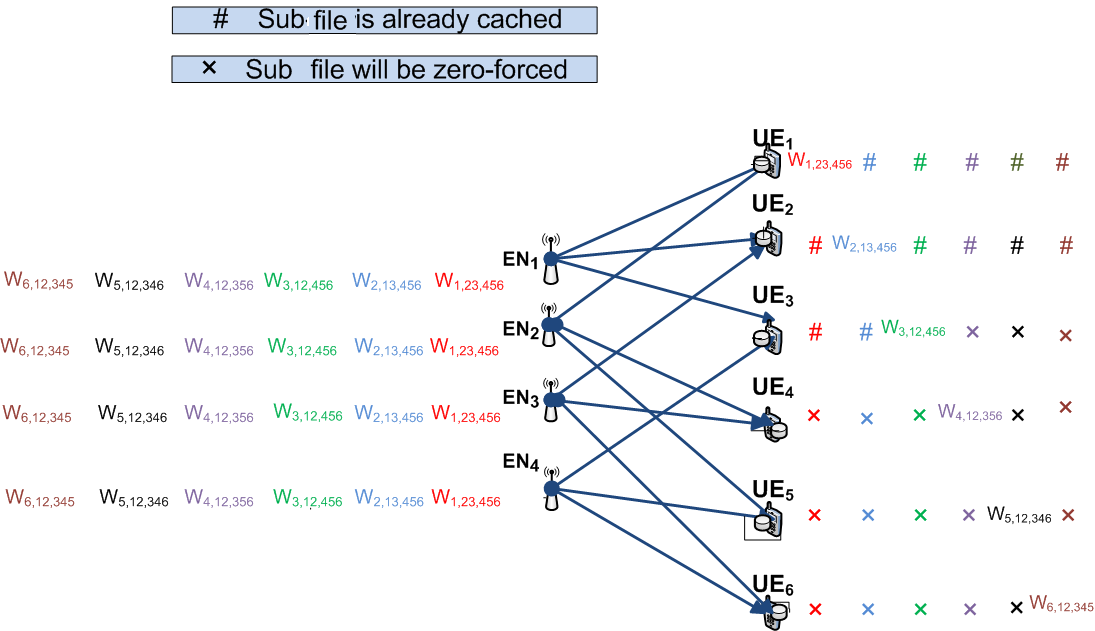}
\caption{ One step of the delivery phase  for the RACN  architecture in Example 2  with receiver connectivity $r=2$, where $H=4$ ENs serve $K = 6$ UEs.}
\label{Example 2}
\end{figure*}

  In Example 2, the delivery phase of the soft-transfer scheme consists of ${6\choose 2}- \binom{5}{1}=10$ steps. At each step all the ENs will cooperate to transmit $K=6$ subfiles, one for each user, by acting  as one transmitter with 4 antennas according to the ideal system assumption. In  Fig. \ref{Example 2}, we show one step of the delivery phase given that all the  ENs  cooperate to deliver the missing subfiles, i.e., the ideal system as  discussed earlier. In this figure, we can see that each user will be able to decode its desired  subfile without any interference, as the undesired subfiles are  either cached or zero-forced at this user. The  edge NDT of this ideal system is $\delta_{E-\text{Ideal}}=\frac{2}{3}$,  while the achievable  end-to-end  NDT   for the soft-transfer scheme is $ \delta_{\text{soft}}(\frac{1}{3},0,\rho)=\frac{2}{3}+\frac{1}{\rho}$.
  \\

 Case 2 ($t_U<K-H$):
  In this case  each subfile  $W_{d_k,\mathcal{T}}$ in \eqref{w}  is partitioned into $\binom{K-(t_U+1)}{H-1}$ disjoints chunks of equal size, denoted by
  \begin{equation}
  \begin{split}
 W_{d_k,\mathcal{T}}=\{W_{d_k,\mathcal{T} , {\pi },{\pi \prime }} :&{\pi } \subseteq[K] \backslash (\{k\} \cup \mathcal{T}),  |\pi|=H-1,\\ & {\pi  \prime} \subseteq[K] \backslash (\{k\} \cup \mathcal{T} \cup  \pi  )\},
 \end{split}
 \end{equation}
 where  $\pi $, $|\pi |=H-1$, is the set of receivers at which  this chunk will be zero-forced, while $\pi  \prime$, $|\pi\prime|=K-(H+t_U)$, is the set of receivers at which  this chunk will cause interference, i.e,   the set of receivers at which  this chunk is  neither cached nor zero-forced.  The  total number of chunks intended for user $\mathrm{UE}_k$ is $\left({K\choose t_U}- \binom{K-1}{t_U-1}\right) \binom{K-(t_U+1)}{H-1} $, while the 
 size of  each chunk is  $|W_{d_k,\mathcal{T} , {\pi },{\pi \prime }}|=\frac{F}{{K\choose t_U}\binom{K-(t_U+1)}{H -1} }$ bits.

 All the  ENs  $\mathcal{E}$ transmit $W_{d_k,\mathcal{T} , {\pi},{\pi \prime}}$ by using the beamforming vector $\mathbf{v}_{\pi }\in \mathbb{R}^{H}$ to zero-force this chunk at the  UEs in set $\pi$.
 The matrix $\mathbf{H}_\pi$ for this case is  of  dimensions $H-1\times H$. The beamforming vector $\mathbf{v}_{\pi } $ is designed as follow
 $ \mathbf{v}_{\pi }=\text{Null}\{\mathbf{H}_{\pi }\},$
 where $\text{Null}\{\mathbf{H}_{\pi }\}$ is the null space of matrix  $\mathbf{H}_{\pi }$. The null space   always exist,  and its size is $1$.

 At each step of delivery phase, we transmit the set of chunks that have the same $\pi \prime$ and belong to different requested files, i.e., we transmit at most one chunk from each requested file at the same time. In other words,  in each step we transmit the set of chunks  that belong to different requested files, but    cause interference at the same set of receivers. The size of each transmitted set per step is $H+t_U$. This is because, the chunks that have the same $\pi \prime$ do  not belong to the set of  requested files 
 $\{W_{d_k} : k \in \pi  \prime \}$,
with size  $ K-(H+t_U)$,  and   will only   belong to  the one of the  remaining $(H+t_U)$   requested files,   denoted by the set
 $ \mathcal{R}= \{W_{d_k} : k \notin \pi  \prime \},$
 while we  only transmit  the chunks that  belong to different  requested files at the same time.
 
  Each user is interested in $\left({K\choose t_U}- \binom{K-1}{t_U-1}\right) \binom{K-(t_U+1)}{H-1} $ chunks, while the total number of requested files is $K$, and in every step we transmit only $(H+t_U)$ chunks. Hence, the total number of  steps  is given by
 \begin{equation}
\frac{\left({K\choose t_U}- \binom{K-1}{t_U-1}\right)\binom{K-(t_U+1)}{H-1}K}{H+t_U}.
 \end{equation}

 We denote the set of UEs  interested in the   transmitted   $(H+t_U)$  chunks that have the same $\pi  \prime $ and belong to different requested files  by 
 $ \mathcal{U}_{\pi  \prime} =\{ \mathcal{U} \backslash  \pi   \prime \} $,
where $|\mathcal{U}_{\pi  \prime}|= H+t_U $.  For       chunk $W_{d_k, \mathcal{T} , \pi, \pi \prime}$    requested by  user $\mathrm{UE}_k \in \mathcal{U}_{\pi  \prime}$, we  have  $(\mathcal{T}  \cup \pi  \cup \pi   \prime )\subseteq \mathcal{U}$; and  hence, $ ( \mathcal{T}  \cup \pi ) \subseteq  \mathcal{U}_{\pi  \prime}$. 
 The set of transmitted chunks will be decoded by the UEs in  set $\mathcal{U}_{\pi  \prime}$ without interference since each chunk $W_{d_k, \mathcal{T} , \pi, \pi \prime}$ in this set is already cached at $|\mathcal{T}|=t_U$ UEs, where $\mathcal{T} \subseteq \mathcal{U}_{\pi  \prime}$, and will be zero-forced at $|\pi |=H-1$  UEs,  $\pi   \subseteq \mathcal{U}_{\pi  \prime}$. Since $ \mathcal{T} \cap \pi  =\phi$ and $|\mathcal{T}|+|\pi|=|\mathcal{U}_{\pi  \prime}|-1$, which means that each chunk will not cause any interference at the $|\mathcal{U}_{\pi  \prime}|-1$ undesired UEs in set $  \mathcal{U}_{\pi  \prime}$. Hence, each user in this set  $\mathcal{U}_{\pi  \prime}$   can decode its desired chunk, the edge NDT of the ideal system  in this case is given by 
 \begin{equation}
 \delta_{E-\text{Ideal}} (\mu_R, \mu_T, \rho) =\frac{K-t_U}{K} \frac{K}{H+t_U}.
 \end{equation}
The achievable  end-to-end  NDT  by using the soft-transfer scheme in case 2 is given by
\begin{equation}
\delta_{\text{soft}}(\mu_R, \mu_T, \rho)  =(K-t_U) \left(\frac{1}{H+t_U}+\frac{1}{ H \rho} \right).
\end{equation}

\subsection{Soft-Transfer Scheme with  Cache Enabled  ENs}\label{subScheme_M}
\subsubsection{\textbf{Cache Placement Phase}}
The  cloud server divides each file $W_n$  in the library, $n\in [N]$,  into two parts $W^1_n$ and $W^2_n$ with sizes $\mu_T F$ bits and  $(1-\mu_T)F$ bits, respectively. 
At first, the cloud server places $W^1_n$, $n\in [N]$,   in the cache memory of all the ENs.   Then,  the UEs apply the  placement scheme in Section \ref{Scheme}-\ref{subScheme}-1  on the two sets     $W^1_n$  and $W^2_n$, $\forall n$.  At the end of the placement phase, each user stores $N \binom {K-1} {t_U-1}$ pieces from each set,  each  piece   $W^1_{n, \mathcal{T}}$, where  $\mathcal{T}\subseteq[K]$, $|\mathcal{T}|=t_U$,  has a size  $\frac{ \mu_T  F}{{K\choose t_U}}$ bits  while the size of  each piece $W^2_{n, \mathcal{T}}$, where  $\mathcal{T}\subseteq[K]$, $|\mathcal{T}|=t_U$,   is   $\frac{ ( 1-\mu_T)    F}{{K\choose t_U}}$, which sum up to $\mu_R NF $ bits, satisfying the memory constraint with equality. 

\subsubsection{\textbf{Delivery  Phase}}
In the first part,  the cloud server  needs to deliver the subfiles 
 \begin{equation}
 \label{w}
 \{W^2_{d_k,\mathcal{T}} : k \notin \mathcal{T}, \mathcal{T} \subseteq[K], |\mathcal{T}|= t_U\}
\end{equation}
to  $\mathrm{UE}_k$, i.e., the subfiles  of file $W^2_{d_k}$ which have not been already stored in the cache of $\mathrm{UE}_k$. The total number of such subfiles is  ${K\choose t_U}- \binom{K-1}{t_U-1}$.  For this case,  we use the same delivery phase  of  the  soft-transfer scheme presented in Section \ref{Scheme}-\ref{subScheme}-2.   The achievable NDT is given by
\begin{equation}
\label{soft-Mt}
\delta_{\text{soft}}(\mu_R, \mu_T, \rho) = (1-\mu_T)(K-t_U) \left[\frac{1}{\min \{H+t_U, K\}}+\frac{1}{ H \rho} \right].
\end{equation}.

In the second part of the delivery phase, the ENs  deliver the subfiles in 
 \begin{equation}
 \label{w_Mx}
 \{W^1_{d_k,\mathcal{T}} : k \notin \mathcal{T}, \mathcal{T} \subseteq[K], |\mathcal{T}|= t_U\}
\end{equation}
to  $\mathrm{UE}_k$, i.e., the subfiles  of file $ W^1_{d_k}$ which have been stored in the cache memory of the ENs and have  not been already stored in the cache of $\mathrm{UE}_k$. The total number of such subfiles is  ${K\choose t_U}- \binom{K-1}{t_U-1}$. For this case,   we can use the same delivery scheme from the ENs to the UEs which is based on ZF and presented in Section \ref{Scheme}-\ref{subScheme}-2,   since each $\mathrm{EN}_i$, $i \in [H]$, caches all  the requested subfiles in \eqref{w_Mx}. According to that,  the end-to-end achievable NDT for the this case  is given by
\begin{equation}
\delta_{\text{soft}}(\mu_R, \mu_T, \rho) = \mu_T(K-t_U) \left  [\frac{1}{\min \{H+t_U, K\}}\right].
\end{equation}

Together with the  achievable NDT in  \eqref{soft-Mt}, we obtain the end-to-end NDT  in Theorem 2.

\section{ Zero-Forcing (ZF) Scheme} \label{zero-forcing}
In this section, we present a centralized coded caching scheme for the same  RACN  architecture with receiver connectivity $r$ with $\mu_R + \mu_T \geq 1$ and $t_R=\frac{(\mu_R+ \mu_T-1)K}{\mu_T}$.
\subsection{Placement Phase}
The  cloud server divides each file $W_n$  in the library, $n\in [N]$,  into two parts $W^1_n$ and $W^2_n$ with sizes $\mu_T F$ bits and  $(1-\mu_T)F$ bits, respectively.  First, the  cloud server places  $W^1_n$, $\forall n$, in the cache memory of all the ENs. 
After that,  each $\mathrm{UE}_k$, $k\in[K]$,  cache the  whole set $\{W^2_n, \; n\in [N]\}$. 
For any file $W^1_n$, $n \in [N]$, we partition it into
 ${K\choose t_R}$ equal-size subfiles, each of  which is denoted by   $W^1_{n, \mathcal{T}}$, where  $\mathcal{T}\subseteq[K]$, $|\mathcal{T}|=t_R$.  The subfiles   $W^1_{n, \mathcal{T}}$, $\forall n$, are stored in the cache memory of  $\mathrm{UE}_k$ if $ k\in \mathcal{T}$.  At the end of the  placement phase, each user stores
$N\binom {K-1} {t_R-1}$ subfiles, each of size $ \frac{ \mu_TF}{{K\choose t_R}}$ bits, from the set of cached subfiles  at the ENs  and stores the set $\{W^2_n, n \in [N]\}$, that has a size of $(1-\mu_T)NF$ bits, which sum up to $\mu_R NF$ bits,  satisfying the cache capacity constraint with equality.
\subsection{Delivery Phase}
 The ENs need to deliver the subfiles in 
 \begin{equation}
 \label{w_M}
 \{W^1_{d_k,\mathcal{T}} : k \notin \mathcal{T}, \mathcal{T} \subseteq[K], |\mathcal{T}|= t_R\}
\end{equation}
to  $\mathrm{UE}_k$, i.e., the subfiles  of file $ W^1_{d_k}$ which have been stored in the cache memory of  the ENs and have  not been already stored in the cache of $\mathrm{UE}_k$. The total number of such subfiles is  ${K\choose t_R}- \binom{K-1}{t_R-1}$. For this case,   we can use  use ZF-based delivery scheme  in Section \ref{Scheme}-\ref{subScheme}-2,   since each $\mathrm{EN}_i$, $i \in [H]$ caches all  the requested subfiles in \eqref{w_M} and hence the ENs can act as one big multi-antenna transmitter where full cooperation is possible among the ENs for any user demand vector. This result in the end-to-end  NDT in Theorem 3.

\begin{figure*}
  \begin{subfigure}[b]{0.49\textwidth}
    \includegraphics[width=\textwidth]{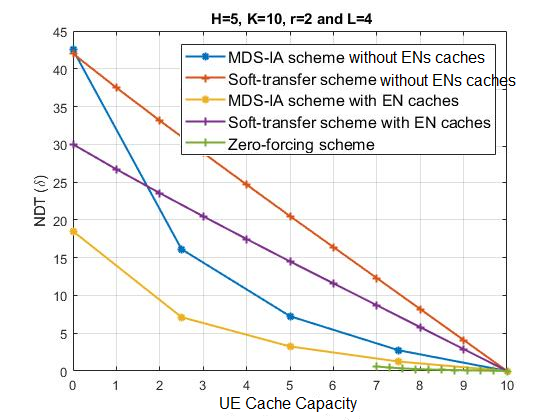}
    \caption{Fronthaul multiplexing gain $\rho=0.05$.}
    \label{fig:1}
  \end{subfigure}
  \begin{subfigure}[b]{0.49\textwidth}
    \includegraphics[width=\textwidth]{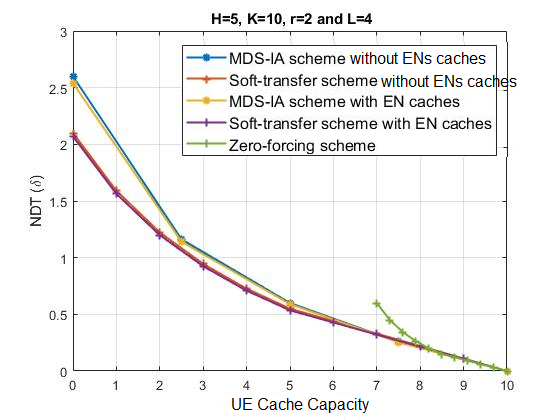}
    \caption{Fronthaul multiplexing gain $\rho=20$.}
    \label{fig:2}
  \end{subfigure}
  \begin{subfigure}[b]{0.49\textwidth}
    \includegraphics[width=\textwidth]{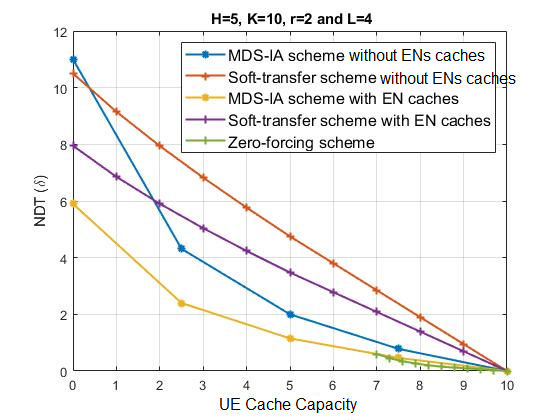}
    \caption{Fronthaul multiplexing gain $\rho=\rho_\text{th}=0.2353$.}
    \label{fig:3}
  \end{subfigure}
   \caption{Comparison of the  achievable NDT for a $5\times10$ RACN  architecture  with library $N=10$ files,  EN's cache size $\mu_T=.3$ and  receiver connectivity $r=2$  for the MDS-IA,  soft-transfer   and  ZF schemes.}
   \label{fig:5}
\end{figure*}

\section{Numerical Results}\label{Numerical Results}
In this section we compare the end-to-end latency achieved by the three  proposed schemes. 
The  NDTs achieved by the MDS-IA,  soft-transfer  and zero-fprcing  schemes  for  three   different fronthaul multiplexing gains are plotted  in Fig. \ref{fig:5}. We observe  that the end-to-end NDT decreases with increasing   $\rho$  and the NDT in Fig. \ref{fig:1} is mainly dominated by the edge NDT for the soft-transfer and the MDS-IA schemes.   We also observe that, caches at the ENs allow reducing the end-to-end NDT as expected, but the amount of reduction becomes negligible as $\rho$ increases, since the files can now be delivered efficiently over the fronthaul links.  We also observe that the best performance among the three schemes depends highly on $\rho$ and the total cache size of the ENs and the UEs. In Fig. \ref{fig:1} and Fig. \ref{fig:3}  with    $\rho=0.05$ and $\rho=0.2353$, respectively, when the total cache size of one UE and one EN is  not sufficient to store the entire library, i.e., $\mu_R < 0.7$, the MDS-IA scheme performs  better than the soft-transfer scheme for almost all cache capacity values, while we observe in Fig. \ref{fig:2} that the soft-transfer scheme  outperforms MDS-IA as the multiplexing gain increases, i.e., $\rho=20$. This is mainly because the edge latency of the soft transfer scheme is minimal as  it is derived based on an ideal fully cooperative delivery from the ENs. Therefore, when the fronthaul links are of high capacity, the performance of the soft transfer scheme becomes nearly optimal. On the other hand, when the fronthaul links are the bottleneck,  latency can be reduced by delivering less information to the ENs, and this is achieved by  MDS-IA.    When the total cache size of one UEs and one EN is   sufficient to store the entire library, i.e., $\mu_R\geq 0.7$, we observe from Fig. \ref{fig:1} that, when the fronthaul multiplexing gain is less than a threshold value $\rho_\text{th}$, i.e., $\rho< 0.2353$, the ZF scheme outperforms others.  This is mainly because, in this scenario the ZF scheme does not use the fronthaul links, which constitute the bottleneck.


\section{Conclusions}\label{Conclusion}
We have studied centralized caching and delivery over a  RACN  with a specified network topology between the ENs and the UEs. We have proposed   three  schemes, namely the MDS-IA, soft-transfer, and ZF schemes. MDS-IA exploits MDS coding for placement and  real IA for delivery. We have derived the achievable NDT for this scheme for an arbitrary cache capacity at the ENs and receiver connectivity of $r=2$, and for an arbitrary  receiver connectivity when the user cache  capacities  are above a certain threshold. The results show that increasing  the receiver connectivity for the same number of ENs and UEs will   reduce the NDT   for a specific cache capacity at the UEs,  while  the amount of reduction depends on the fronthaul multiplexing gain, $r$. 
We also consider the soft-transfer scheme which  quantizes and transmits   coded symbols to  each of the ENs  over the fronthaul links, in order to implement  ZF over the edge network. Finally, the ZF scheme  is presented for an arbitrary value of $r$ when the total cache size at one EN and one UE is sufficient to store the whole library. This allows all user requests to be
satisfied by ZF from the ENs to the UEs without the participation of the cloud server.  We have observed  that  when the total cache size of the UE and the EN is not sufficient to store the entire library, the  MDS-IA scheme  performs  better  when the fronthaul multiplexing gain is limited, while the soft-transfer scheme outperforms MDS-IA when the fronthaul multiplexing gain is high. On the other hand, when the total cache
size of one UE and one EN is sufficient to store the entire library and the fronthaul capacity is below a certain threshold, ZF  achieves the smallest NDT.


\nocite{*}
\bibliographystyle{IEEEtran}
\bibliography{IEEEabrv,mybibfile}
\end{document}